\documentclass[11pt]{article}
\usepackage{amsmath, amssymb, amscd, amsthm, amsfonts}
\usepackage{graphicx}
\usepackage{hyperref}
\usepackage{algorithm}
\usepackage{algpseudocode}
\usepackage{caption}
\usepackage{algorithmicx}
\usepackage{float}
\usepackage{indentfirst}
\usepackage[T1]{fontenc}
\usepackage[utf8]{inputenc}
\usepackage{authblk}
\newcommand\keywords[1]{\textbf{Keywords}: #1}

\title{Feasibility Analysis of Grover-meets-Simon Algorithm}
\author[1,2,3]{Qianru Zhu}
\author[4]{Huiqin Xie}
\author[1,2,3]{Qiqing Xia}
\author[1,2,3]{Li Yang  \thanks{Corresponding author: yangli@iie.ac.cn}}
\affil[1]{State Key Laboratory of Information Security, Institute of Information Engineering, Chinese Academy of Sciences,Beijing, China}
\affil[2]{Institute of Information Engineering,Chinese Academy of Sciences,Beijing, China}
\affil[3]{School of Cyber Security,University of Chinese Academy of Sciences, Beijing, China}
\affil[4]{Department of cryptography and Technology, Beijing Electronic Science and Technology Institute, Beijing, China}
\date{}

\begin{document}

\maketitle

\begin{abstract}
Quantum algorithm is a key tool for cryptanalysis. At present, people are committed to building powerful quantum algorithms and tapping the potential of quantum algorithms, so as to further analyze the security of cryptographic algorithms under quantum computing. Recombining classical quantum algorithms is one of the current ideas to construct quantum algorithms. However, they can not be easily combined, the feasibility of quantum algorithms needs further analysis in quantum environment.

This paper reanalyzes the existing combined algorithm——Grover-meets-Simon algorithm in terms of the principle of deferred measurement. First of all, due to the collapse problem caused by the measurement, we negate the measurement process of Simon's algorithm during the process of the Grover-meets-Simon algorithm. Second, since the output of the unmeasured Simon algorithm is quantum linear systems of equations, we discuss the solution of quantum linear systems of equations and find it feasible to consider the deferred measurement of the parallel Simon algorithm alone. Finally, since the Grover-meets-Simon algorithm involves an iterative problem, we reconsider the feasibility of the algorithm when placing multiple measurements at the end. According to the maximum probability of success and query times, we get that the Grover-meets-Simon algorithm is not an effective attack algorithm when putting the measurement process of the Simon algorithm in the iterative process at the end of Grover-meets-Simon algorithm.
\end{abstract}


\section{Introduction}
% Since the Shor algorithm\cite{shor1994algorithms}  and Grover algorithm\cite{grover1996fast} were proposed, quantum computing has attracted the attention in cryptography. Shor's algorithm posed a serious threat on the security of cryptographic algorithms based on large number prime factorization and discrete logarithm problems. Grover algorithm can speed up the key search. 
With the development of quantum computers, the threat of quantum computing to cryptographic algorithms cannot be ignored. Currently, people are fully tapping the potential of quantum algorithms to attack cryptographic algorithms. To effectively analyze the security of current cryptographic algorithms under quantum computing. This will help people better design cryptographic algorithms to resist attacks from future quantum computers.

Quantum algorithms are the main research objects in the field of cryptanalysis. 
The attack on the public key system originated from the fact that Shor's algorithm\cite{shor1994algorithms} transformed the factorization problem into an order problem. Shor's algorithm can accelerate the large integer decomposition problem exponentially from the quantum perspective. Later, based on adiabatic quantum computation\cite{farhi2001quantum}, new thought was put forward on the decomposition of large integers. In 2001, Burges\cite{burges2002factoring} transformed the integer factorization problem into an optimization problem for the first time, laying the foundation for the application of adiabatic quantum computing to integer factorization. Later, Schaller and Schutzhold\cite{schaller2007role} improved the method. The integer factorization problem based on adiabatic quantum theory is mainly divided into two aspects. One is the study of integer factorization of pairs of NMR quantum processors\cite{dattani2014quantum}\cite{pal2019hybrid}\cite{peng2008quantum}\cite{xu2012quantum}. The second is the research based on the D-wave quantum annealing algorithm.\cite{finnila1994quantum}\cite{dridi2017prime}\cite{peng2019factoring}\cite{warren2019factoring}.
In symmetric cryptanalysis, the application and promotion of Grover's algorithm is the most widely used quantum algorithm in this field. The Grover algorithm can speed up the search efficiency of unordered data sets and reduce the complexity of exhaustive search attacks from $O(2^n)$ to $O(\sqrt{2^n})$. Therefore, people consider extending the key length of the symmetric cipher to resist the attack of Grover's algorithm. Then, Brassard et al. generalize the Grover algorithm and get the current QAA algorithm\cite{brassard2002quantum}. Compared to Grover's algorithm, the QAA algorithm extends the preparation of the initial state from the $n$ Hadamard gates to any quantum algorithm $\mathcal{A}$, thus expanding the application range. Based on the Grover algorithm, Brassard et al.\cite{brassard1998quantum} proposed a quantum collision search algorithm (BHT algorithm) for the 2-to-1 function, which can find a set of collisions with high probability through $O(2^{\frac{n}{3}})$ queries. Ambainis\cite{ambainis2007quantum} extends the scope of application to arbitrary functions and finds a set of collisions with $O(2^{\frac{2n}{3}})$ queries. Furthermore, Aaronson and Shi\cite{aaronson2004quantum} proved that the lower complexity limit of the BHT algorithm is $O(2^{\frac{n}{3}})$, which is lower than the original best lower limit. And Zhandry\cite{zhandry2013note} proves the probability that the probability of finding a collision after performing q queries is $\frac{q^3}{2^n}$ at most. Furthermore, Hosoyamada et al.\cite{harrow2009quantum} explored multicollisions and proposed a new quantum algorithm. For $l$-collisions, when $l$ is small, $2^{\frac{n\times(3^{l-1}-1)}{2\times3^{l-1}}}$ queries get collisions. In addition to Grover's algorithm, Simon's algorithm\cite{simon1997power} is also the research focus of current cryptanalytic quantum algorithms. Kuwakado and Morii\cite{kuwakado2010quantum} proved that a three-round Feistel structure distinguisher was constructed through Simon's algorithm using the parallelism characteristics of quantum computers. The three-round Feistel structure and random permutation distinction, which is not possible for the classical computer. Similarly, many cryptanalytic works (\cite{kuwakado2012security}\cite{santoli2017using}\cite{kaplan2016breaking}) are carried out based on the feature that the Simon algorithm can find the period of Boolean functions. Many common quantum attack methods(\cite{ito2019quantum}\cite{xie2018quantum}\cite{roetteler2015note}\cite{bonnetain2020quantum}) are also the application and extension of Simon's algorithm.

In recent years, the focus of many research work is to propose new quantum attack algorithms for cryptographic algorithms. Many people try to combine different quantum algorithms to obtain a more efficient algorithm, such as Simon meets Kuperberg algorithm, Grover meets Kuperberg algorithm\cite{bonnetain2018hidden}, Bernstein–Vazirani meet Grover algorithm\cite{zhou2021quantum}, etc. The most widely used is the Grover-meets-Simon algorithm proposed by Leander and May\cite{leander2017grover}. Grover's algorithm is used as the outer structure, and Simon's algorithm is used as the inner structure. The period found by Simon's algorithm is used as the judgment condition of Grover's search. With the introduction of the Grover-meets-Simon algorithm, people began to consider using this algorithm to search for two keys for the function structure that can construct a period function\cite{dong2019quantum}\cite{dong2018quantum}\cite{shinagawa2022quantum}\cite{grover1996fast}. Bonnetain\cite{bonnetain2019quantum} improved it and proposed that the combined algorithm of Grover and Simon under the Q1 model. It can be used to improve the quantum query complexity by using the "offline calculation + online query" method. 
%However, most cryptographic attack quantum algorithms are based on idealized quantum computing models without considering the effect of entanglement.
However, the probability of success of Grover's algorithm depends on various situations. This paper mainly discusses the Grover-meets-Simon algorithm in terms of the deferred measurement.

\subsection*{Our contributions} 

1.We propose the concept of quantum linear systems of equations. By constructing the quantum Gaussian elimination algorithm, we discuss the situation of solving linear systems of equations in superposition state(quantum linear systems of equations) based on the quantum Gaussian elimination algorithm. On the basis of solving quantum linear systems of equations, we found that the parallel Simon algorithm is feasible to defer the measurement. This is the first time to consider the feasibility of quantum Gaussian elimination algorithm in quantum linear systems of equations in detail. 

2. We re-analyze the probability of success and query times of Grover-meets-Simon algorithm in terms of the principle of deferred measurement. First, we consider that when Simon's algorithm is used as the inner structure, if the measurement is assumed, the key space will collapse. Hence, when Simon's algorithm is not measured, that is, deferred measurement after Grover-meets-Simon algorithm, we consider the problem that moving all Simon's algorithm measurements to the end of the entire algorithm during the iteration will cause. Because there is an oscillation in the initial state, the probability of success and the query times of Grover-meets-Simon algorithm are re-analyzed to explore whether it is an effective quantum attack algorithm in the quantum environment. Ensuring the real feasibility of quantum attack algorithm in quantum environment is one of the problems that should be considered when constructing quantum attack algorithm. This paper provides a new idea for the real feasibility analysis of quantum algorithms.

\subsection*{Outline} 

The remainder of the paper is arranged as follows. In section 2, we introduce the knowledge of linear algebra and quantum attack algorithms. In section 3, We analyze the initial state prepared by the Grover-meets-Simon algorithm and Simon's measurement problem. In section 4, we discuss the situation of solution when quantum Gaussian elimination algorithm solve linear systems of equations in superposition state(quantum linear systems of equations). In section 5, we reanalyze the feasibility of Grover-meets-Simon algorithm. In section 6, we give a summary of this paper.

\section{Preliminaries }
\subsection*{Related contents of linear algebra}

In this section, we will introduce some basic knowledge about solving linear systems of equations. Given a matrix $A$ and a vector $\vec{b}$, find a vector $\vec{x}$ so that $A\vec{x}=\vec{b}$.  Among classical algorithms for solving linear systems of equations, the most basic one is Gaussian elimination algorithm.

 \begin{algorithm}[H]
		\caption{\textbf{Gaussian elimination algorithm}}
		\label{Alg:GEA}
		\begin{algorithmic}[1]
			\Require 
		    $[A|b]$:Augmented matrix belongs to $C^{n\times(n+1)}$, where $A$ is a $n\times n$ matrix and $rank(A)=n$
			\Ensure  the value of a vector $\vec{x}$ such at $A\vec{x}=\vec{b}$
			\State \textbf{for} $j\leftarrow 1$ $\textbf{to}$ $n-1$  
			\State \ \ \ \ $\textbf{for} \ \ \ i\leftarrow j+1 $     $\textbf{to}$ $n$
			\State \ \ \ \ \ \ \ \    $c=-a_{i,j}/a_{j,j}$
			\State \ \ \ \ \ \ \ \    \textbf{for} $k\leftarrow 1$ $\textbf{to}$ $n$
			\State \ \ \ \ \ \ \ \ \ \ \ \  $a_{i,k}=a_{i,k}+a_{j,k}\times c$ 
			\State \ \ \ \ \ \ \ \    $b_i=b_i+b_j\times c$
			\State $x_n=b_n/a_{n,n}$
			\State \textbf{for} $j\leftarrow n-1$ $\textbf{to}$ $1$ 
			\State \ \ \ \ $\textbf{for} \ \ \ i\leftarrow j+1 $     $\textbf{to}$ $n$	
			\State \ \ \ \ \ \ \ \    $b_j=b_j-a_{j,i}\times x_i$
			\State \ \ \ \ $x_{j}=b_{j}/a_{j,j}$
			\State \textbf{return} $x$
		\end{algorithmic}
\end{algorithm}

When exploring the principle of Gaussian elimination, the following two matrices are often involved:

\textbf{Definition 1 (Row echelon form matrix): }If a matrix is  row echelon form matrix, it satisfies the following conditions:

(1)If it has both zero and non-zero rows, the zero row is below and the non-zero row is above.

(2) If it has nonzero rows, the column number of the first nonzero element of each nonzero row is strictly monotonically increasing from top to bottom.
\[
\begin{bmatrix} a_{1,1} & a_{1,2}& a_{1,3} & \dots &a_{1,n-2}&a_{1,n-1}&a_{1,n}
\\0 & 0& a_{2,3} & \dots &a_{2,n-2}&a_{2,n-1}&a_{2,n}\\
0 & 0& 0 & \dots &a_{3,n-2}&a_{3,n-1}&a_{3,n}\\
\vdots & \vdots&\vdots& \ddots&\vdots&\vdots&\vdots\\
0 & 0& 0 & \dots &0&0&a_{m,n}\\
0 & 0& 0 & \dots &0&0&0\\
\vdots & \vdots&\vdots& \ddots&\vdots&\vdots&\vdots\\
0 & 0& 0 & \dots &0&0&0
\end{bmatrix} .
\]

\textbf{Definition 2 (Row simplest form matrix): }In a row echelon form matrix, if the first non-zero element of a non-zero row is 1, and the other elements in the column are all zero, the row echelon form matrix is called the row simplest matrix.
\[
\begin{bmatrix} 1 & a_{1,2}& 0 & \dots &a_{1,n-2}&0&a_{1,n}
\\0 & 0& 1 & \dots &a_{2,n-2}&0&a_{2,n}\\
0 & 0& 0 & \dots &a_{3,n-2}&0&a_{3,n}\\
\vdots & \vdots&\vdots& \ddots&\vdots&\vdots&\vdots\\
0 & 0& 0 & \dots &0&1&a_{m,n}\\
0 & 0& 0 & \dots &0&0&0\\
\vdots & \vdots&\vdots& \ddots&\vdots&\vdots&\vdots\\
0 & 0& 0 & \dots &0&0&0
\end{bmatrix}
\]

Among them, the row echelon form matrix is the intermediate matrix form after the elimination step of the Gaussian elimination method. And the row simplest form matrix is the final matrix form after the operation of the Gauss Jordan elimination method. In the section 4,%改章节
we will use the idea of solving the basic solution system in linear algebra. After the elementary transformation of the augmented matrix, we will get a row simplest matrix, as follows:
\begin{equation}\label{eqexpmuts}
\begin{aligned}[b]
    [A|b]\stackrel{elementary\ \  transformation}{\longrightarrow}\begin{bmatrix} 1 & \dots & 0 &a'_{1,r+1}&\dots& a'_{1,n} & b'_1\\ 
\vdots &\ddots & \vdots & \vdots & \vdots& \vdots& \vdots \\
0 & \dots & 1 & a'_{r,r+1} &\dots & a'_{r,n} & b'_r\\
0 & \dots& 0 & 0 &\dots& 0 & 0 \\
\vdots & \ddots & \vdots & \vdots & \vdots &\vdots & \vdots\\
0 & \dots& 0 & 0 &\dots& 0 & 0 
\end{bmatrix}
\end{aligned}
\end{equation}

According to linear algebra, we can know that when $rank(A)=r$, the number of basis vectors of the basic solution system is $n-r$, and they have the following form:
\[
\eta_1=\begin{bmatrix}-a'_{1,r+1}\\ \vdots \\ -a'_{r,r+1}\\ 1\\0\\ \vdots \\ 0\end{bmatrix},\eta_2=\begin{bmatrix}-a'_{1,r+2}\\ \vdots \\ -a'_{r,r+2}\\ 0\\1\\ \vdots \\ 0\end{bmatrix},\dots ,\eta_{n-r}=\begin{bmatrix}-a'_{1,n}\\ \vdots \\ -a'_{r,n}\\ 0\\0\\ \vdots \\ 1\end{bmatrix}
\]
The special solution of linear systems of equations is $x_0=[b'_1,\dots, b'_r,0,\dots,0]^T$. We can get the general solutions of linear systems of equations $x=x_0+k_1\eta_1+\dots+k_{n-r}\eta_{n-r}$, where $k_1,\dots,k_{n-r}$ are arbitrary constants.

\subsection*{Quantum circuit}
In this section, we briefly introduce the relevant knowledge of quantum circuits. The circuit composed of multiple quantum gates with certain logic function is called quantum circuit.It is composed of wires for transmitting information and quantum gates for processing information, which can be used to describe the changes of quantum states. Each quantum logic gate can be represented by a unitary matrix. Quantum gates acting on $n$ qubits can be represented by $2^n\times 2^n$ unitary matrix. 

A single qubit has two quantum ground states $|0\rangle$,$|1\rangle$. If a qubit is in a state $|\varphi\rangle$ other than ground states and can be expressed linearly by $|0\rangle$,$|1\rangle$, this state is called superposition state  $|\varphi\rangle=\alpha|0\rangle+\beta|1\rangle$. The probability amplitude $\alpha$ and $\beta$ is plural and meets $|\alpha|^2+|\beta|^2=1$.
Common single qubit gates and their matrices are shown in figure 1. In this paper, the X gate is frequently used to realize qubit inversion, such as $X|c\rangle=|1\oplus c\rangle$. And $X$ represents the operation of Pauli-X gate.

\begin{figure}[H]
\centerline{\includegraphics[width=10cm]{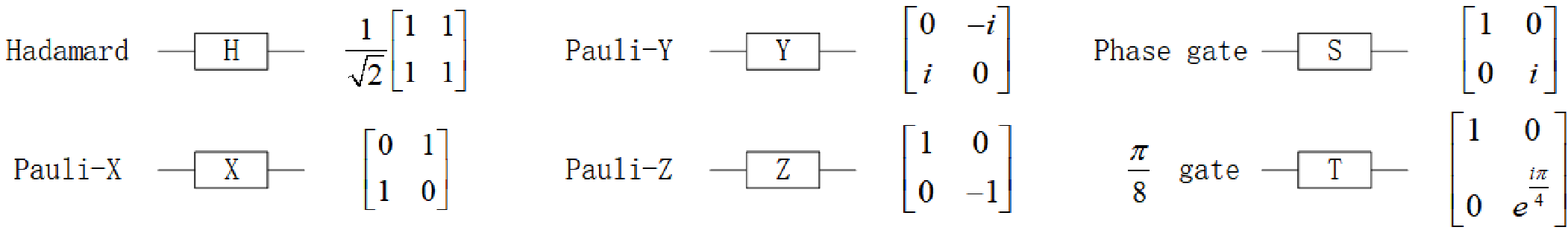}}  
  \caption{Common single qubit gates}
\end{figure}

In the quantum circuit, multiple-qubit gates are used frequently. Common multiple-qubit gates include CNOT gate, Toffoli gate, SWAP gate, and so on, as shown in figure 2. XOR gates in classical circuits can be realized by CNOT gates in quantum logic gates. Similarly, Toffoli gates can realize the 'AND' operation in quantum computing, and Toffoli gates can also be regarded as controlled CNOT gates. In ion trap quantum computers, CNOT gates can only be operated serially. Even if different CNOT gates involve different qubits, they cannot be operated in parallel\cite{yang2013post}. Therefore, the number of CNOT gates greatly affects the running time of quantum algorithms. When we consider quantum algorithms, we focus on the number of CNOT gates. Toffoli gate is a commonly used quantum gate and needs to be decomposed into Clifford + T gates for running.  In this paper, we adopt the decomposition method in \cite{nielsen2002quantum}. That is, a Toffoli gate can be decomposed into seven T gates, six CNOT gates, two H gates, and one S gate. It can be seen that six CNOT gates can be reduced by reducing one Toffoli gate. In our paper, we mainly focus on the reduction of toffoli gates. Besides a Toffoli gate is with 7 T-depth, which is one of the focuses of quantum circuit cost.

\begin{figure}[H]
\centerline{\includegraphics[width=8cm]{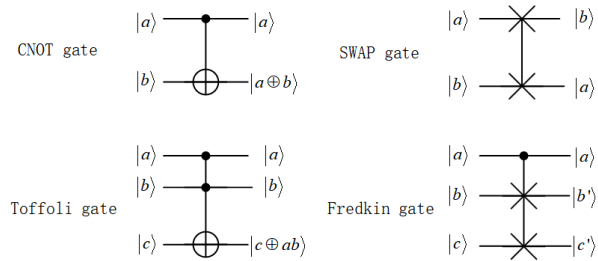}} 
  \caption{Common multiple-qubit gates}
\end{figure}

When constructing quantum Gaussian elimination algorithm, it will involve many multi-controlled gates. They can be decomposed into a series of Toffoli gates\cite{nielsen2002quantum}. As shown in figure 3, when we decompose a n-fold controlled-U gates, the circuit can be divided into three parts: $n$ control qubits, $n-1$ auxiliary qubits and a controlled qubit. Therefore, we can obtain $2(n-1)$ Toffoli gates and a controlled-U gate.

\begin{figure}[h!]
\centerline{ \includegraphics[width=10cm]{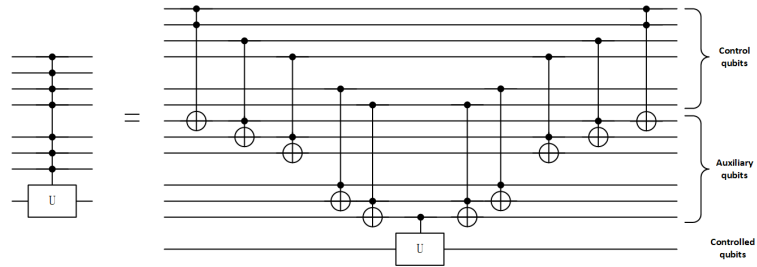}} 
  \caption{The decomposition of n-fold controlled-NOT}
\label{fig:decompose}
\end{figure}

\subsection*{Quantum algorithm}
\subsubsection*{Simon's algorithm}
Simon's algorithm \cite{simon1997power} mainly solves the period finding problem in cryptanalysis:

\textbf{Problem 1(Simon's problem)}Let $f$ : $\{0,1\}^n\rightarrow X$ be a function such that for all $x,y\in F_2^n$ with $x\ne y$, $f(x)=f(y)\Leftrightarrow x=y\oplus s$. Given a oracle to $f$ to find $s$.

Simon's algorithm like this:

1.Initialize 2n qubits  $|0^n\rangle|0^n\rangle$, and apply H gates on the first n qubits, we can get

$$\frac{1}{\sqrt{2^n}}\sum_{x\in F_2^n}|x\rangle|0^n\rangle$$

2. Apply $O_f$ on upper state, we can get
$$\frac{1}{\sqrt{2^n}}\sum_{x\in F_2^n}|x\rangle|f(x)\rangle$$

3.Apply H gates on the first n qubits, we can get
$$\frac{1}{2^n}\sum_{x\in F_2^n}\sum_{y\in F_2^n}(-1)^{x\cdot y}|y\rangle|f(x)\rangle$$

4.If $f(x)$ satisfies that $f(x)=f(x\oplus s)$, the state can be write that
$$\frac{1}{2^n}\sum_{x\in X_1}\sum_{y\in F_2^n}((-1)^{x\cdot y}+(-1)^{(x\oplus s)\cdot y})|y\rangle|f(x)\rangle$$ 
where $X_1$ is the $n-1$ dimensional subspace of $F_2^n$, and divides $F_2^n$ into coset $X_1$ and $X_1+s$. 

5.Measure the register and return them.

From Simon's algorithm, we can see when $(-1)^{x\cdot y}+(-1)^{(x\oplus s)\cdot y}\ne 0$, $y$ will satisfy $y\bot s$ after measurement. Hence, we can only get the vector $y$ orthogonal to period $s$. Repeat the above steps $O(n)$ times, we can get $n-1$ vectors orthogonal to $s$ with high probability. Then we can solve the linear systems of equations to get $s$.

\begin{figure}[h!]
\centerline{ \includegraphics[width=10cm]{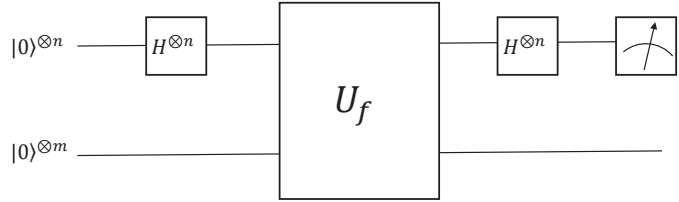}} 
  \caption{Simon's algorithm circuit}
\end{figure}

\subsubsection*{Grover's algorithm}
Grover's algorithm\cite{grover1996fast} is a quantum search algorithm which aims to find marked elements in set X. Compared with classical search algorithms, Grover's algorithm brings quadratic speed-up. And Theorem 1 give a generalized version of Grover's algorithm, QAA algorithm\cite{brassard2002quantum},

\textbf{Theorem 1 : } Let $\mathcal{A}$ be any quantum algorithm without measurement, and let $g$: $F_2^n\to F_2$ be any Boolean function that distinguish between good and bad states from output of algorithm $\mathcal{A}$. Let $p>0$ be the initial success probability of
$\mathcal{A}|0\rangle$. We define $\mathcal{Q=-A}$$U_0$$\mathcal{A}U_g$, where $U_0=2|0\rangle\langle0|-I$, $U_g$ is that if the state is good, phase reversal is performed:
$$U_g|x\rangle=\begin{cases}
-|x\rangle,\ \ if\ \  g(x)=1\\
\ \ |x\rangle, \ \ if\ \  g(x)=0
\end{cases}$$
when we compute $\mathcal{Q}^m\mathcal{A}|0\rangle$ and measure the system, the outcome is good with probability at least max $\{p,1-p\}$, where $m=[\frac{\pi}{4\theta}]$ and $sin^2(\theta)=p$.

\begin{figure}[H]
\centerline{ \includegraphics[width=10cm]{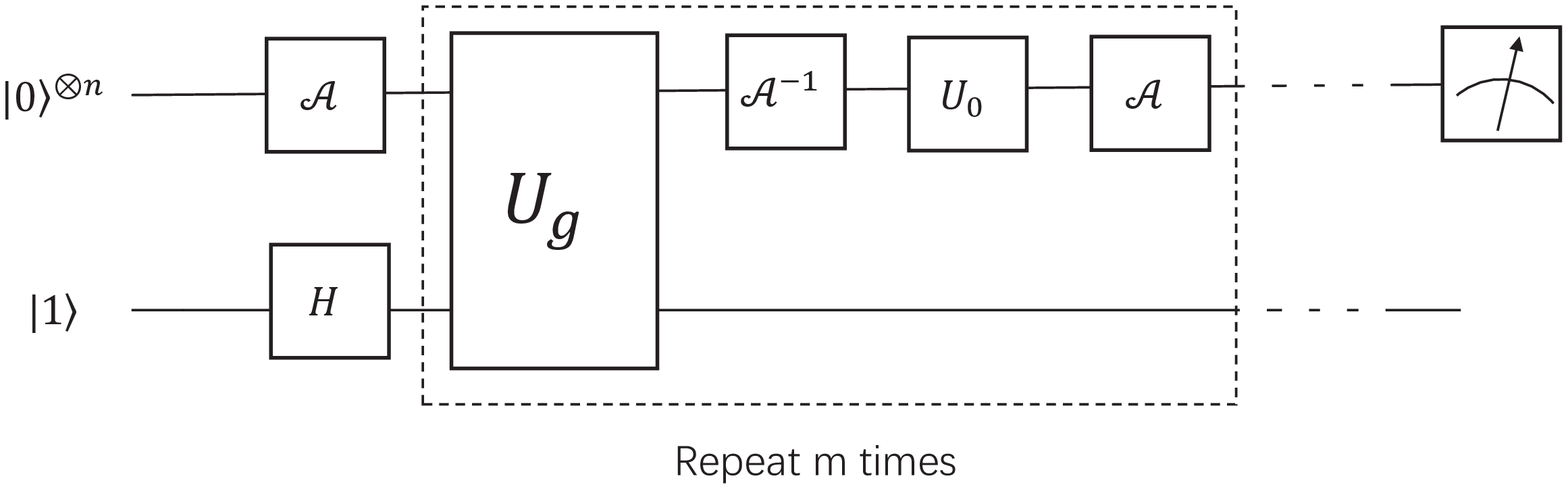}} 
  \caption{Grover's algorithm circuit}
\end{figure}

\section{The initial state of Grover-meets-Simon algorithm}

Grover-meets-Simon algorithm, as a quantum attack algorithm for symmetric cryptography such as FX-construction, is often used to search for two unknown keys. In the FX construction, $Enc(x)=E(k,x+k_1)+k_2$. Suppose that $f(k',x)= Enc(x)+E(k',x)$, obviously we can know when $k'=k$, $f(k',x)$ have the period $s=k_1$. 

\begin{figure}[h!]
\centerline{ \includegraphics[width=10cm]{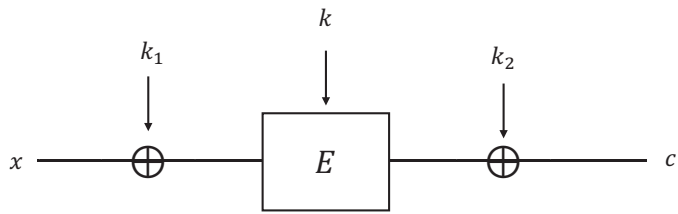}} 
  \caption{FX-construction}
\end{figure}

We combine the algorithms of Grover and Simon. Grover's algorithm is to search the key $k$. Simon's algorithm is a classifier to judge whether the function $f(k',x)$ has a period $k_1$. In order to judge the actual feasibility of Grover-meets-Simon algorithm running in a quantum environment, we need to analyze its entire running process. 

% \begin{figure}[h!]
% \centerline{ \includegraphics[width=8cm]{A.eps}} 
%   \caption{Initial state preparation of Grover-meets-Simon algorithm}
% \end{figure}

\textbf{Lemma 1:} Let $f: \{0,1\}^m\times \{0,1\}^n\rightarrow\{0,1\}^n$ be a function such that $f(k',x)=Enc(x)+E(k',x)$. Let us apply algorithm $\mathcal{A}$ in \cite{leander2017grover} on $|0\rangle^{\otimes m+2nl}$, we can obtain the superposition:
\begin{equation}
    \label{equ1}
\begin{aligned}[b]
 |\varphi\rangle=&\frac{1}{\sqrt{2^m}}(\frac{1}{2^n})^l\sum_{k'\in F_2^m,k'\neq k}\sum_{x_1,\dots x_l\in F_2^n}\sum_{y_1,\dots y_l\in F_2^n}(-1)^{x_1\cdot y_1}\dots(-1)^{x_l\cdot y_l}|y_1,y_2\dots y_l\rangle|f(k',x_1),\dots,f(k',x_l)\rangle|k'\rangle\\
+&\frac{1}{\sqrt{2^m}}(\frac{1}{2^{n-1}})^l\sum_{x_1 \dots x_l\in X_1}\sum_{y_1,\dots y_l\perp s,s\neq 0}(-1)^{x_1\cdot y_1}\dots(-1)^{x_l\cdot y_l}|y_1,y_2\dots y_l\rangle|f(k,x_1),\dots,f(k,x_l)\rangle|k\rangle
\end{aligned}
\end{equation}

\textbf{Proof:} After apply algorithm $\mathcal{A}$ in \cite{leander2017grover}, the initial state is

$$|\varphi\rangle=\frac{1}{\sqrt{2^m}}(\frac{1}{2^n})^l\sum_{k'\in F_2^m}\sum_{x_1,\dots x_l\in F_2^n}\sum_{y_1,\dots y_l\in F_2^n}(-1)^{x_1\cdot y_1}\dots(-1)^{x_l\cdot y_l}|y_1,y_2\dots y_l\rangle|f(k',x_1),\dots,f(k',x_l)\rangle|k'\rangle$$

Considering a Simon's algorithm alone, we can get

\begin{equation}\label{equ2}
\begin{aligned}[b]
|0\rangle^{\otimes n}|0\rangle^{\otimes n}\stackrel{H^{\otimes n}\otimes I}{\longrightarrow}&\frac{1}{2^n}\sum_{x\in F_2^n}|x\rangle|0\rangle^{\otimes n}\stackrel{U_{f_{k'}}}{\longrightarrow}\frac{1}{\sqrt{2^n}}\sum_{x\in F_2^n}|x\rangle|f(k',x)\rangle\\
=&\frac{1}{\sqrt{2^n}}\sum_{x\in F_2^n}\frac{1}{2}(|x\rangle+|x\oplus s\rangle)|f(k',x)\rangle\\
\stackrel{H^{\otimes n}\otimes I}{\longrightarrow}&\frac{1}{\sqrt{2^n}}\frac{1}{\sqrt{2^n}}\sum_{x\in F_2^n}\sum_{y\in F_2^n}\frac{1}{2}(-1)^{x\cdot y}(1+(-1)^{y\cdot s})|y\rangle|f(k',x)\rangle \\
=&\frac{1}{\sqrt{2^n}}\frac{1}{\sqrt{2^n}}\sum_{x\in F_2^n}\sum_{y\in F_2^n,y\perp s}(-1)^{x\cdot y}|y\rangle|f(k',x)\rangle.
\end{aligned}
\end{equation}

Consider that if $k$ is the correct key, then the function $f(k,x)$ has a non-zero period $s$ such that $f(k,x)=f(k,x\oplus s)$, where $x\in F_2^n$. To simplify the representation, we define a set $X_1$ , where $X_1$ is the $n-1$ dimensional subspace of $F_2^n$, and divides $F_2^n$ into coset $X_1$ and $X_1+s$. We suppose that the ground states in formula (\ref{equ2}) can be combined to obtain this result.

\begin{equation}\label{k'=k}
\begin{aligned}[b]
\frac{1}{\sqrt{2^{n-1}}}\frac{1}{\sqrt{2^{n-1}}}\sum_{x\in X_1}\sum_{y\in F_2^n,y\perp s}(-1)^{x\cdot y}|y\rangle|f(k,x)\rangle
\end{aligned}
\end{equation}

For the function $f(k',x)$, suppose that $k$ is the correct key. If $k'=k$, the period $s$ is not 0; if $i\neq k$, the period $s$ is 0. Then the initial state of $\mathcal{A}$ algorithm is $|\varphi\rangle=\mathcal{A}|0\rangle$, that is
 \begin{align*}
 |\varphi\rangle=&\frac{1}{\sqrt{2^m}}(\frac{1}{2^n})^l\sum_{k'\in F_2^m,k'\neq k}\sum_{x_1,\dots x_l\in X_1}\sum_{y_1,\dots y_l\in F_2^n}(-1)^{x_1\cdot y_1}\dots(-1)^{x_l\cdot y_l}|y_1,y_2\dots y_l\rangle|f(k',x_1),\dots,f(k',x_l)\rangle|k'\rangle\\
+&\frac{1}{\sqrt{2^m}}(\frac{1}{2^{n-1}})^l\sum_{x_1 \dots x_l\in X_1}\sum_{y_1,\dots y_l\perp s,s\neq 0}(-1)^{x_1\cdot y_1}\dots(-1)^{x_l\cdot y_l}|y_1,y_2\dots y_l\rangle|f(k,x_1),\dots,f(k,x_l)\rangle|k\rangle
 \end{align*}
 \hfill$\qedsymbol$

Next, we reanalyze the initial state $|\varphi\rangle$. In formula (\ref{equ1}), the key space $\sum_{k'\in F_2^n}|k'\rangle$ to be searched is entangled with the superposition state of the matrices $\sum_{y_1,y_2\dots y_l\in F_2^n}|y_1,y_2\dots y_l\rangle$ generated in paralle multiple Simon's registers. When $|k'\rangle$ is the correct key $|k\rangle$, the row vector $|y_i\rangle(i\in {1,2\dots l})$ of the corresponding matrix $|y_1,y_2\dots y_l\rangle$ satisfies that $y_i\in \{y|y\bot s,s\ne 0\}$. When $|k'\rangle$ is not the correct key $|k\rangle$, the row vector $|y_i\rangle(i\in {1,2\dots l})$ of corresponding matrix $|y_1,y_2\dots y_l\rangle$ satisfies $y_i\in F_2^n$. Suppose that $U_1=\{(y_1,y_2\dots y_l)\in F_2^{nl}|y_i\bot s,s\ne 0\} $, $U_2=\{(y_1,y_2\dots y_l)\in F_2^{nl}|y_i\in F_2^n \} $. Obviously, $U_1\subsetneqq U_2$. If we consider the measurement in the original Simon's algorithm, the key space $\sum_{k'\in F_2^n}|k'\rangle$ will collapse so that we can not search the correct key $|k\rangle$. Therefore, we can only measure at the end of the algorithm.

In addition, when $O(n)$ Simon's algorithm operate in parallel, the linear systems of equations generated is different from the linear systems of equations generated by a single Simon register running $O(n)$ times. Due to the measurement process, after $O(n)$ times Simon's algorithm, we can obtain a determined matrix. When the number of operations of Simon's algorithm increases, $dim(span((y_1,y_2\dots y_l)))=n-1$ can be obtained with a high probability\cite{kaplan2016breaking}, so that the period $s$ can be obtained. However, without measurement after $O(n)$ Simon's registers operating in parallel, the superposition state of matrices $\sum_{y_1,y_2\dots y_l\in F_2^n}|y_1,y_2\dots y_l\rangle$ would will contain all matrices generated by a row vector y that satisfies the condition. That is, the matrices components with $rank < n-1$ will always exist.  For example, one of the components of the superposition state is $|y_1,y_1\dots y_1\rangle$, that is all row vectors are same and $dim((y_1,y_1\dots y_1))=1$. Therefore, we not only search for the correct key $k$, but also search for the superposition components $|y_1,\dots,y_l\rangle$ which can obtain the period $s$ by Gaussian elimination. 

First, because the measurement of the Simon algorithm is postponed to the end of the entire algorithm, we must consider the specific process and impact of solving the solution of the superposition state of linear systems of equations $\sum_i\alpha_i|y_1,y_2\dots y_l\rangle$ in subsequent calculations.   

\section{Quantum Gaussian Elimination Algorithm for Solving Quantum Linear Systems of Equations }

\subsection{Solving Linear Systems of Equations in Non Superposition State}

In order to consider the solution of the linear systems of equations in the superposition state, we first consider the solution method of the non-superposition state. In this section, we consider the quantum implementation of the classical algorithm for solving a linear systems of equations. It should be noted here that the coefficient matrix of linear equations in non superposition state mentioned in this algorithm is full rank.

Similar to the classical Gaussian elimination algorithm, we also divide the quantum Gaussian elimination algorithm into two parts. The first part is called forward elimination, which  reduces a given system to row echelon form. The second part is called back substitution, which puts the matrix into reduced row echelon form.  

\begin{figure}[h!]
\centerline{  \includegraphics[width=10cm]{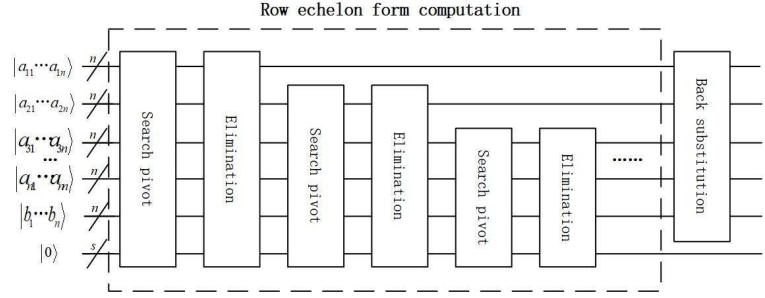}}

  \caption{Quantum Gaussian elimination algorithm}
  \label{figqge}
\end{figure}

Algorithm \ref{Alg:REFC} gives a description in pseudo-code on how to construct the row echelon form. The algorithm in this paper is similar to the idea in \cite{perriello2021complete}, and is optimized on some toffoli gates. In algorithm \ref{Alg:REFC}, $a_{j,:}$ represents elements of the $j-th$ row of the matrix $A$. Given a matrix $A$ and a vector $\vec{b}$, find a vector $\vec{x}$ such at $A\vec{x}=\vec{b}$. When $A$ is a $n\times n$ invertible matrix, algorithm 2 can be changed to a correct upper triangular matrix B. In the language of quantum state, suppose that $|A\rangle$ represents the quantum state of $A$, that is, $|A\rangle=|a_{1,1},a_{1,2},\dots, a_{1,n},a_{2,1},\dots a_{2,n},\dots ,a_{n,n}\rangle$. 

\begin{algorithm}
		\caption{\textbf{Row echelon form computation}}
		\label{Alg:REFC}
		\begin{algorithmic}[1]
			\Require 
		    $m\times n$ qubit to store the matrix value of $A$:$|a_{1,1},a_{1,2}\dots,a_{1,n},a_{2,1},\dots a_{2,n},\dots a_{m,n}\rangle$, $O(n^2)$ auxiliary qubits to ensure that the algorithm runs.
			\Ensure matrix  $B$ corresponding to $A$.
			% \State $\textbf{if}$ $n<m$ $\textbf{do}$
			\State \ \ \ \ \textbf{for} $j\leftarrow 1$ $\textbf{to}$ $n$  
			\State \ \ \ \ \ \ \ \  $\textbf{for} \ \ \ i\leftarrow j+1 $     $\textbf{to}$ $m$
			\Comment{pivot exchange}
			\State \ \ \ \ \ \ \ \ \ \ \ \  $\textbf{if}$  $a_{j,j}=0$   $\textbf{do}$  $a_{j,:}=a_{j,:}\bigoplus a_{i,:}$ 
			\State \ \ \ \ \ \ \ \ $\textbf{for} \ \ \ k\leftarrow j+1 $     $\textbf{to}$ $m$
			\Comment{elimination}
			\State \ \ \ \ \ \ \ \ \ \ \ \  $\textbf{if}$  $a_{k,j}=1$   $\textbf{do}$ $a_{k,:}=a_{j,:}\bigoplus a_{k,:}$
			% \State $\textbf{else}$ $n\ge m$ $\textbf{do}$
			% \State \ \ \ \ \textbf{for} $j\leftarrow 1$ $\textbf{to}$ $m-1$  
			% \State \ \ \ \ \ \ \ \  $\textbf{for} \ \ \ i\leftarrow j+1 $     $\textbf{to}$ $m$
			% \Comment{pivot exchange}
			% \State \ \ \ \ \ \ \ \ \ \ \ \  $\textbf{if}$  $a_{j,j}=0$   $\textbf{do}$  $a_{j,:}=a_{j,:}\bigoplus a_{i,:}$ 
			% \State \ \ \ \ \ \ \ \ $\textbf{for} \ \ \ k\leftarrow j+1 $     $\textbf{to}$ $m$
			% \Comment{elimination}
			% \State \ \ \ \ \ \ \ \ \ \ \ \  $\textbf{if}$  $a_{k,j}=1$   $\textbf{do}$ $a_{k,:}=a_{j,:}\bigoplus a_{k,:}$		
			\State \textbf{return} $B=A$
		\end{algorithmic}
\end{algorithm}

Using algorithm \ref{Alg:REFC} in algorithm \ref{Alg:FE}, we can obtain a complete quantum Gaussian elimination algorithm to solve linear systems of equations.

 \begin{algorithm}[H]
		\caption{\textbf{Quantum Gaussian elimination algorithm}}
		\label{Alg:FE}
		\begin{algorithmic}[1]
			\Require 
		    $n(n+1)$ qubits to store the augmented matrix value of $[A|b]$:$|a_{1,1},a_{1,2}\dots,a_{1,n},a_{2,1},\dots a_{2,n},\dots a_{n,n},b_1,b_2\dots b_n\rangle$, $O(n^2)$ auxiliary qubits to ensure that the algorithm
			\Ensure the value of a vector $\vec{x}$ such at $A\vec{x}=\vec{b}$
			\State Run algorithm 2
			\State \textbf{for} $j\leftarrow n$ $\textbf{to}$ $2$ \Comment{Back substitution solution}
			\State \ \ \ \ $\textbf{for} \ \ \ i\leftarrow j-1$ $\textbf{to}$ $1$
			\State \ \ \ \ \ \ \ \   $\textbf{if}$  $a_{i,j}=1$ $\textbf{do}$ $b_{i}=b_{j}\bigoplus b_{i}$
			\State \textbf{return} $x=b$
		\end{algorithmic}
\end{algorithm}

Classical Gaussian elimination algorithm is mostly applied to dense matrices, so there is a necessary condition that diagonal elements are not 0. In the quantum Gaussian elimination algorithm, we consider that the matrix elements belong to $F_2$. Therefore, the probability that the diagonal elements are 0 is very high. To ensure that the diagonal elements are not 0, we need to use row exchange to determine the primary element.

When we run the algorithm \ref{Alg:REFC}, we divide the whole process into two parts: determine the pivot element and eliminate the element, as shown in Figure \ref{figqge}. 

\begin{figure}[h!]
\centerline{  \includegraphics[width=10cm]{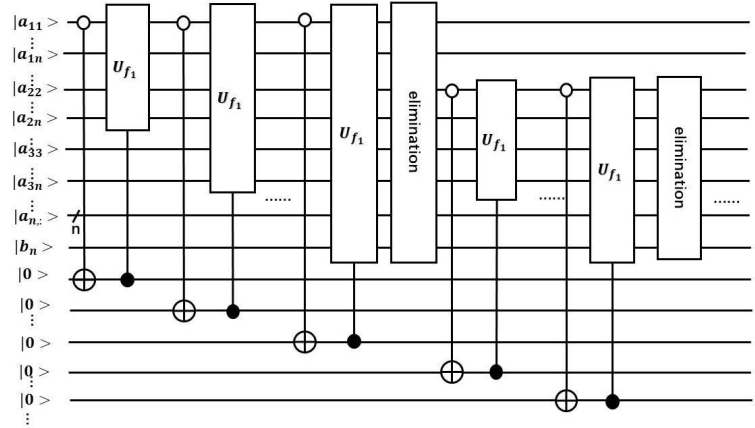}}

  \caption{The process of searching pivots}
  \label{pivot}
\end{figure}

In figure \ref{pivot}, we show that the detailed quantum circuit is used to search the pivot process. When we search for the pivot of $i-th$ column, let the function $f_1$ be: 
\[
f_1(a_{i,j},a_{k,j})=(a_{i,j}\oplus a_{k,j},a_{k,j}) \ \ \ \forall j\ge i \ and\ \ k>i .
\]
Operator $U_{f_1}$ can be realized by some CNOT gates. Before the operation $U_{f_1}$ in every row, we all need store the value of $a_{i,i}$ in auxiliary qubits firstly. At the same time, since operation $U_{f_1}$ is controlled by an auxiliary qubit, it can be seen that the process of exchanging rows of matrix can be realized by  some Toffoli gates. In addition, we need to add auxiliary qubits to record the change of the value of the pivot. Consider that searching the pivot of $i-th$ column, we need $n-i$ CNOT gates, $(n-i)(n-i+2)$ Toffoli gates and $n-i$ auxiliary qubits

\begin{figure}[h!]
\centerline{ \includegraphics[width=10cm]{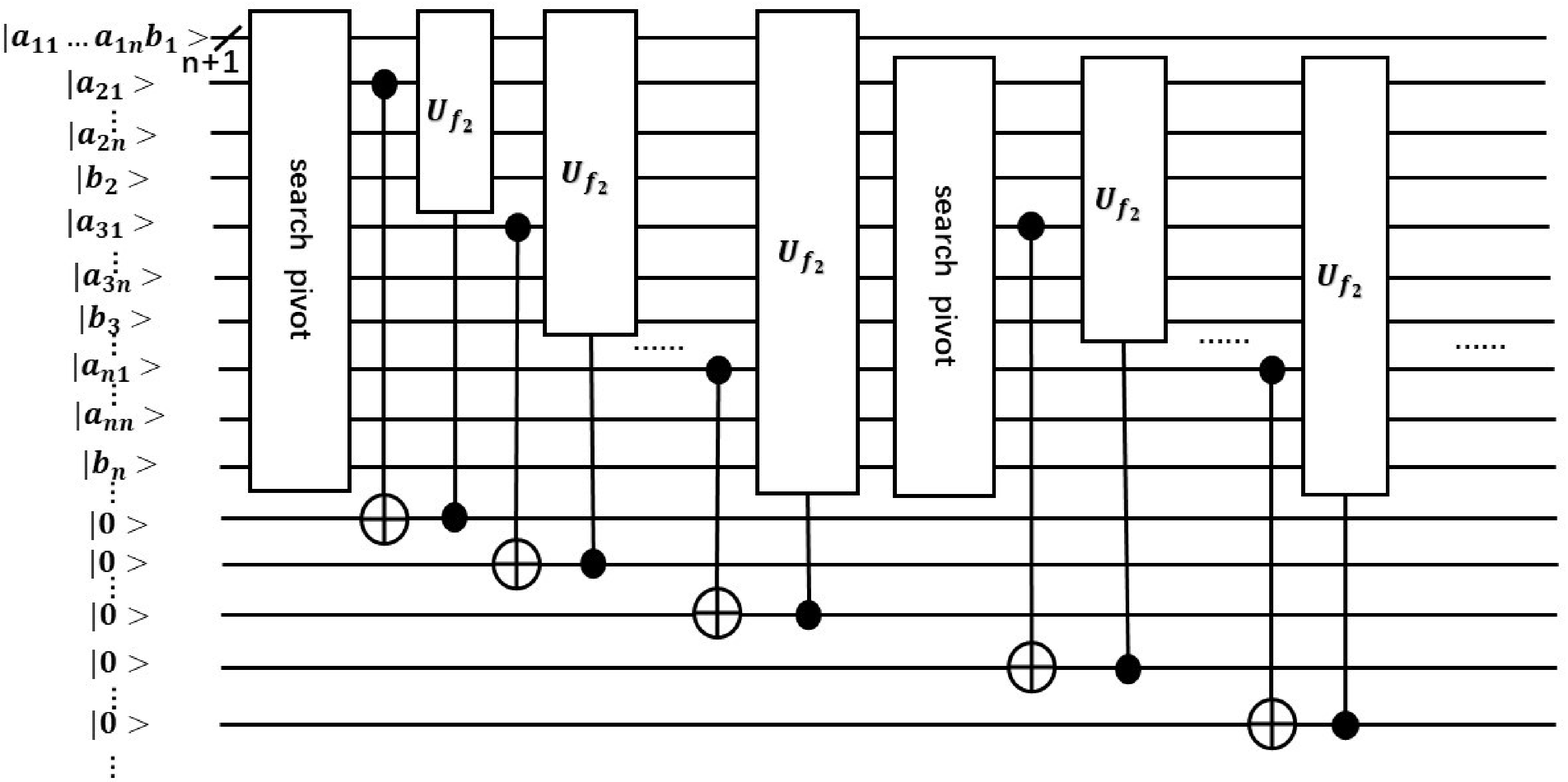}}

  \caption{The process of elimination}
  \label{elimination}
\end{figure}

For another process-elimination operation, as shown in figure \ref{elimination}. When we eliminate the $i-th$ column, let function $f_2$ be: 
\[
f_2(a_{i,j},a_{k,j})=(a_{i,j}, a_{i,j}\oplus a_{k,j}) \ \ \forall j\ge i\ \ and\ \ k>i .
\]

Like function $f_1$, $f_2$ can also be realized by CNOT gates, then the controlled $U_{f_2}$ can be realized by Toffoli gates.  Auxiliary qubits storage the value of $a_{k,i}(k>i)$. Then auxiliary qubits will control the operation of function $f_2$, which is equivalent to controlling the operation of function $f_2$ according to the value of  $a_{k,i}(k>i)$. Obviously, when $a_{k,i}(k>i)=1$, the quantum circuit can complete the elimination operation. Here, we can simply optimize the the number of toffoli gates. For the $i$ th column $a_{k,i}(k>j)$, we can just use the CNOT gates to change its value to 0. To sum up, to eliminate the $i-th$ column, we need $2(n-i)$ CNOT gates , $(n-i)(n-i+1)$ Toffoli gates and $n-i$ auxiliary qubits.

In order to obtain the solution of quantum linear systems of equations, we also need to consider the quantum circuit implementation of the back substitution process. According to algorithm 3, we know that Toffoli gates can perform this process. As shown in figure \ref{BSS}, $\frac{n(n-1)}{2}$ Toffoli gates are needed. According to article\cite{yang2013post}, CNOT gates  affect the operation speed of the quantum algorithm to a certain extent. When computing quantum resources, we decompose Toffoli gates into CNOT gates and T-depth for consideration. Quantum Gaussian elimination algorithm needs $4n^3-\frac{15}{2}n^2-\frac{23}{2}n$ CNOT gates. And T-depth is $\frac{7n(n-1)(2n+5)}{3}$, auxiliary qubits are $n(n-1)$.

\begin{figure}[h!]
 \centerline{  \includegraphics[width=8cm]{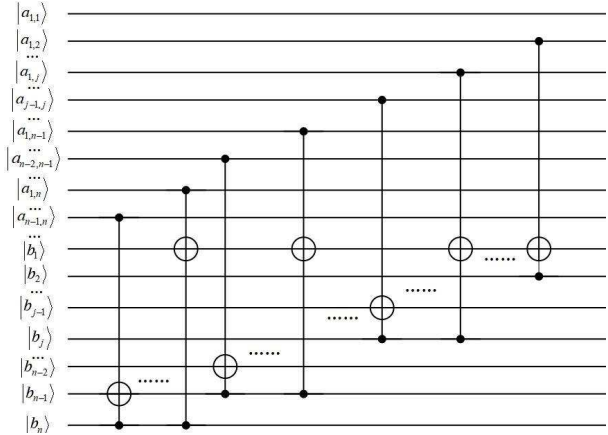}}
  \caption{The process of back substitution solution}
  \label{BSS}
\end{figure}

In Appendix, we also give the quantum Gauss-Jordan algorithm. And based on the quantum Gaussian Jordan elimination algorithm, we also give a method to solve the row simplest form matrix of the non-full rank matrix.

\subsection{Solving Linear Systems of Equations in Superposition State}

It can be obtained from section 3 that we need to analyze the solution of linear systems of equations when it is input in a superposition state. Compared with non superposition state input, superposition state input is more common in quantum algorithms. Next, we consider the quantum Gaussian elimination algorithm in the input of the superposition state. First, to facilitate discussion, we give a definition. Note that to find the solution of the linear systems of equations or the rank of linear systems of equations, the Gaussian elimination algorithm needs to consider the transformation of the row simplest form matrix first. Therefore, this section mainly discusses the problems when the row simplest form matrix is obtained.

\textbf{Definition 3: (Quantum linear systems of equations): } Similar to classical linear systems of equations, when the coefficient matrix of linear systems of equations is superposition, that is $|A\rangle=\sum_i\alpha_i|A_i\rangle$, we call it a quantum linear systems of equations.

% \textbf{Definition 4: (The solution of quantum system of linear equations): } Given a quantum system of linear equations $|A\rangle=\sum_i\alpha_i|A_i\rangle$, its solution is defined as the superposition of the corresponding solution of each linear system $\sum_j\beta_j|x_j\rangle$

% Obviously, if we can get the solution of quantum system of linear equations, like $\sum_j\beta_j|s_j\rangle$, it will change the amplitude of the initial state of Grover-meets-Simon algorithm and then we will get different iterations. 

According to the circuit discussion in  section 4.1, quantum Gaussian elimination algorithm can guarantee that the input is a superposition state. If we only consider the transformation to the row-minimum matrix transformation, we can construct the quantum algorithm $U_{QGE(R)}$:
$$U_{QGE(R)}(\sum_i\alpha_i|A_i\rangle|0\rangle)=\sum_i\alpha_i|A_i\rangle|A'_i\rangle$$
where $A'_i$ is the row simplest form matrix corresponding to $A_i$.

\begin{figure}[H]
 \centerline{  \includegraphics[width=10cm]{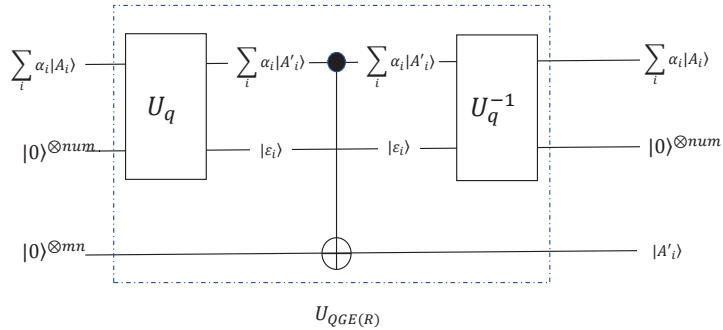}}
  \caption{Quantum Circuit Construction of $U_{QGE(R)}$}
  \label{BSS}
\end{figure}

Different row simplest matrix corresponds to different solution. Therefore, there is always a algorithm to obtain the solution from the row simplest matrix. Suppose there is a algorithm $U_{QGE}$ such that $U_{QGE}(\sum_i\alpha_i|A_i\rangle|0\rangle)=\sum_i\alpha_i|A_i\rangle|s_i\rangle$,where $s_i$ is a solution corresponding to $A_i$. 

In addition, due to the complexity of quantum entanglement, we fully consider another situation  quantum  linear systems of equations, whether quantum Gaussian elimination algorithm can realize $\sum_i\alpha_i|A_i\rangle\rightarrow\sum_j\beta_j|B_j\rangle$. Therefore, it is necessary to consider the feasibility of solving the quantum linear systems of equations from the perspective of auxiliary qubits.  

\textbf{Lemma 2:} Auxiliary qubits $|0\rangle^{\otimes num}$ are necessary to complete the quantum Gaussian elimination algorithm. It is equivalent that suppose that there is a operation $U_{QGE}$ representing quantum Gaussian elimination algorithm such that
\[
U_{QGE}(|A\rangle|b\rangle|0\rangle^{\otimes num})=|B\rangle|x\rangle|\varepsilon\rangle^{\otimes num},
\]
where $num$ is the number of auxiliary qubits, $|B\rangle$ is the upper triangular matrix corresponding to $|A\rangle$, and $|\varepsilon\rangle$ is auxiliary qubits after quantum Gaussian elimination algorithm. Then it holds that $num>0$.

\textbf{Proof: } First, we need prove that in the quantum Gaussian elimination algorithm, the controlled gate is necessary. According to the principle of traditional Gaussian elimination algorithm, that is, when the augmented matrix $[A|b]$ becomes $[C|d]$ through elementary row transformation, the solutions of $Ax=b$ and $Cx=d$ are the same. So we can use elementary row transformation to transform the augmented matrix into row simplest form matrix, and then find the solution of the linear systems of equations. The execution of the elementary row transformation in the two operations of exchanging the pivot and elimination is controlled by the element in the column where the primary element is located. For example, to complete the operation of the $i-th$ column, we need to change $|a_{k,i}\rangle$ to $|1\rangle$, where $|a_{k,i}\rangle$ is the pivot of the $i-th$ column. Searching the $i-th$ column from $k-th$ row to bottom, if $|a_{j,i}\rangle = |1\rangle(j\ge k)$, the row will be exchanged to the $i-th$ row through row transformation.  Obviously, the condition for the row transformation to operate is $(a_{k,i}=0)\wedge(a_{j,i}=1)(j>k)$. Therefore, the controlled gate is necessary in quantum Gaussian elimination algorithm.

Second, we need to consider the construction of controlled gates in quantum Gaussian elimination algorithm. Only consider the pivot $|a_{k,i}\rangle$ and a certain element $|a_{j,i}\rangle(j>k)$ of its column. According to the previous analysis, these two elements is not only as control qubits, but also need to be transformed as target qubits when performing pivot search. So suppose there exists a transformation $U$ such that $U|00\rangle=|00\rangle$,$U|01\rangle=|10\rangle$,$U|10\rangle=|10\rangle$,$U|11\rangle=|11\rangle$. From this, we can find that 
% %
% \begin{equation}\label{eqexpmuts}
% \begin{aligned}[b]
% \begin{cases}
% U|01\rangle=|10\rangle\\
% U|10\rangle=|10\rangle\\
% \end{cases}
% =
% \begin{cases}

% \end{cases}
% \end{aligned}
% \end{equation}
% %

$$
\left\{ 
\begin{matrix}
U|01\rangle=|10\rangle\\
U|10\rangle=|10\rangle\\
\end{matrix}
\right.
\Rightarrow\left\{ 
\begin{matrix}
U(0,0,1,0)^T=(0,1,0,0)^T\ \ (i)\\
U(0,1,0,0)^T=(0,1,0,0)^T\ \ (ii)\\
\end{matrix}
\right.
$$
If we do $(i)-(ii)$, we get that $U(0,-1,1,0)^T=(0,0,0,0)^T$. There is no inverse matrix $U^{-1}$ such that $U^{-1}(0,0,0,0)^T=(0,-1,1,0)^T$. $U$ is not a unitary matrix, so the suppose does not hold. 

It can be seen from the foregoing that we need add auxiliary qubits to store the value of data qubits to control the transformation. \hfill$\qedsymbol$

\textit{$ Remark: $ If the quantum Gaussian elimination method does not require the output of the upper triangular matrix, there will be another situation. The essence of Gaussian elimination algorithm is a process of transforming columns in order through row transformation. That is to say, after the elimination process of column $|a_{1,j},a_{2,j}\dots a_{m,j}\rangle$ is completed, there will be no secondary operations on the column elements. Therefore, if only to find the solution of the linear systems of equations, column elements $|a_{1,j},a_{2,j}\dots a_{m,j}\rangle$ can be used as control qubits, and the last $n-j$ column elements of the matrix can be used as target qubits for transformation. Since the quantum computer cannot label the elements in sequence, when determining the pivot, it is necessary to find the first element that is not 0 through multi-controlled gates.The decomposition of multi-controlled gates will inevitably require auxiliary qubits\cite{nielsen2002quantum}.}

From lemma 2, we know that it is necessary to add auxiliary qubits in quantum Gaussian elimination algorithm. If we want to get the superposition state of the solution of linear systems of equations, we should consider that whether auxiliary qubits can be disentangled.

\textbf{Theorem 2: }There is no operator $U$, where $U=U_{QGE}\cdot U_2$,  such that
$$\sum_{i}\alpha_i|A_i\otimes0^{num}\rangle\stackrel{U}{\longrightarrow}(\sum_j\beta_j|B_j\rangle)\otimes(\sum_k\gamma_k|c_k\rangle) $$
where $|A_i\rangle$ is the quantum state representation of the matrix, and $|B_j\rangle$ is the quantum state representation of the row simplest form matrix corresponding to $|A_i\rangle$.

\textbf{Proof: }Suppose that there is an operator $U$ such that $\sum_{i}\alpha_i|A_i\otimes0^{num}\rangle\stackrel{U}{\longrightarrow}(\sum_j\beta_j|B_j\rangle)\otimes(\sum_k\gamma_k|c_k\rangle) $. We divide $U$ into the product of two unitary operators $U_{QGE}$ and $U_2$. According to theorem 2, 

\begin{equation}\label{equ:t2}
\begin{aligned}
    U_{QGE}((\sum_{i}\alpha_i|A_i\rangle)\otimes|0\rangle^{\otimes num}) &= \sum_{i}\alpha_i U_{QGE}(|A_i\rangle\otimes|0\rangle^{\otimes num})\\
&=\sum_{i}\alpha_i(|{A_i}'\rangle\otimes|\varepsilon_i\rangle)\\
&\stackrel{U_2}{\longrightarrow}(\sum_j\beta_j|B_j\rangle)\otimes(\sum_k\gamma_k|c_k\rangle) \\
&=\sum_{j,k}\beta_j\gamma_k|B_j\otimes c_k\rangle
\end{aligned}  
\end{equation}

where $A'_i$ is the row simplest form matrix corresponding to $A_i$. 
Therefore, we convert the original proposition to whether auxiliary qubits added in the operation can be disentangled through $U_2$. Since $U_2$ is a unitary operator, the dimension of the space where $\sum_i\alpha_i(|A_i'\rangle\otimes|\varepsilon_i\rangle)$ is located is equal to the dimension of the space where $\sum_{j,k}\beta_j\gamma_k|B_j\otimes c_k\rangle$ is located. 

Case 1: $\sum_j=1$ and $\sum_k\ne 1$
% $<|\varepsilon_1\rangle,|\varepsilon_2\rangle\dots|\varepsilon_i\rangle>=<|c_1\rangle,|c_2\rangle\dots|c_k\rangle>$.

Because of auxiliary qubits record the information of $|A_i\rangle$. When $i\ne i'$, $|A_i\rangle\ne |A_{i'}\rangle$. We get $j=1$. That is, for different $|A_i\rangle$, we must get the same row simplest form matrix $|B\rangle$. Obviously, the case does not hold.

Case 2: $\sum_j\ne1$ and $\sum_k= 1$

According to formula (\ref{equ:t2}), when $k=1$, that is, auxiliary qubits return to all zero states. We can get that $\sum_i\alpha_i|A_i\rangle\stackrel{U}{\longrightarrow}\sum_j\beta_j|B_j\rangle$.
% % Suppose that $A_i$ is an arbitrary $m\times n$ matrix. Hence, The space where $|A\rangle$ is located is the space generated by the matrix based on $|A_i\rangle$. The space where $B_j$ is located is the space generated by all row simplest form matrices. 
% Different matrices may get the same row simplest form matrix after quantum Gaussian elimination algorithm, that is $|\{A_i\}|>|\{B_j\}|$.
Suppose that $|B_1\rangle$ is the row simplest form matrix corresponding to $|A_{11}\rangle$, $|A_{12}\rangle$. And $|B_2\rangle$ is the row simplest form matrix corresponding to $|A_{21}\rangle$, $|A_{22}\rangle$. There are superposition states $\frac{1}{\sqrt{2}}|A_{11}\rangle+\frac{1}{\sqrt{2}}|A_{21}\rangle$ and $\frac{1}{\sqrt{2}}|A_{12}\rangle+\frac{1}{\sqrt{2}}|A_{22}\rangle$. We get that
$$U(\frac{1}{\sqrt{2}}|A_{11}\rangle+\frac{1}{\sqrt{2}}|A_{21}\rangle)=U(\frac{1}{\sqrt{2}}|A_{12}\rangle+\frac{1}{\sqrt{2}}|A_{22}\rangle)=\frac{1}{\sqrt{2}}|B_{1}\rangle+\frac{1}{\sqrt{2}}|B_{2}\rangle$$
Consider the inverse operation of $U$, $U^{-1}(\frac{1}{\sqrt{2}}|B_{1}\rangle+\frac{1}{\sqrt{2}}|B_{2}\rangle)$ has a unique value. Contradicting with the above formula, this case does not hold.

% we can combine the same row simplest matrix in $|A'_i\rangle$, and $dim(\sum_j\beta_j|B_j\rangle)<dim(\sum_i\alpha_i|A_i\rangle)$. Unitary transformations cannot change the dimension of space, so this case does not hold.

Case 3: $\sum_j\ne1$ and $\sum_k\ne 1$

In this case, we can get $\sum_{i}\alpha_i|A_i\otimes0^{num}\rangle\stackrel{U}{\longrightarrow}(\sum_j\beta_j|B_j\rangle)\otimes(\sum_k\gamma_k|c_k\rangle) $.  Suppose that there are $k_1$ original matrices $|A_{i_1}\rangle$ corresponding to the same row simplest form matrix $|B_1\rangle$ and  there are $k_2$ original matrices $|A_{i_2}\rangle$ corresponding to the same row simplest form matrix $|B_2\rangle$. 

\begin{equation}\label{eqeth2}
\begin{aligned}[b]
\begin{cases}
U_{QGE}(|A_{i_1}\otimes0^{num}\rangle)=|B_1\otimes\varepsilon_{i_1}\rangle,\ \ \ i_1\in \{1,2,\dots k_1\}\\
U_{QGE}(|A_{i_2}\otimes0^{num}\rangle)=|B_2\otimes\varepsilon_{i_2}\rangle,\ \ \ i_2\in \{1,2,\dots k_2\}
\end{cases}.
\end{aligned}
\end{equation}

Obviously, the number of original matrices corresponding to different row simplest form matrices is different. Hence, $\sum_i\alpha_i(|A_i'\rangle\otimes|\varepsilon_i\rangle)\ne\sum_{j,k}\beta_j\gamma_k|B_j\otimes c_k\rangle$

Besides, if $\sum_{i}\alpha_i|A_i\otimes0^{num}\rangle\stackrel{U}{\longrightarrow}(\sum_j\beta_j|B_j\rangle)\otimes(\sum_k\gamma_k|c_k\rangle)$, then there is always a unitary transformation that changes auxiliary bits $(\sum_k\gamma_k|c_k\rangle)$ into all-zero states. As in case 2, this case does not hold.

 In summary, there is no unitary transformation $U_2$, equivalent to this theorem, which does not hold.\hfill$\qedsymbol$

According to the above analysis, quantum Gaussian elimination algorithm can realize $\sum_i\alpha_i|A_i\rangle\rightarrow\sum_j\beta_j|B_j\rangle$.  

% Due to the nature of quantum entanglement, we have no way to know the specific situation of the quantum register where the solution of system of linear equations is located, so it will affect the subsequent operations.

% \textit{$ Remark: $ In this article, we consider that we change data qubits representing matrix to get the row simplest form matrix. Hence, auxiliary qubits would record the information of the original matrix. If we use auxiliary qubits to record the row simplest form matrix, the information of the original matrix would be recorded in data qubits as original input, like $U_{QGE}(\sum_i\alpha_i(|A_i\rangle|0^{num}\rangle))=\sum_i\alpha_i(|A_i\rangle|s_i\rangle)$. Both cases are equivalent for the entanglement of auxiliary qubits and data qubits. }

Based on quantum algorithm $U_{QGE}$ such that $U_{QGE}(\sum_i\alpha_i|A_i\rangle|0\rangle)=\sum_i\alpha_i|A_i\rangle|s_i\rangle$, we discuss the deferred measurement of parallel multiple Simon algorithms. 

When measuring $l$ Simon's registers before solving linear systems of equations, for each register we get some $y_i$ vector perpendicular to period $s$ with probability $\frac{1}{2^{n-1}}$. Then for $l$ registers, the number of matrices we finally get with $y_i$ as the row vector is $(\frac{1}{2^{n-1}})^l$. Suppose that the number of matrices with $rank=n-1$ is $r$. Then the probability of getting the correct period is $\frac{r}{2^{(n-1)l}}$. 

If we compute quantum linear systems of equations before measurement, according to formula (\ref{k'=k}), we can get that
$$(\frac{1}{2^{n-1}})^l\sum_{x_1 \dots x_l\in X_1}\sum_{y_1,\dots y_l\perp s,s\neq 0}(-1)^{x_1\cdot y_1}\dots(-1)^{x_l\cdot y_l}|y_1,y_2\dots y_l\rangle|f(x_1),\dots,f(x_l)\rangle$$
After solving the quantum linear systems of equations, we get that
$$(\frac{1}{2^{n-1}})^l\sum_{x_1 \dots x_l\in X_1}\sum_{y_1,\dots y_l\perp s,s\neq 0}(-1)^{x_1\cdot y_1}\dots(-1)^{x_l\cdot y_l}|y_1,y_2\dots y_l\rangle|s_{(y_1,y_2\dots y_l)}\rangle|f(x_1),\dots,f(x_l)\rangle$$
where $|s_{(y_1,y_2\dots y_l)\rangle}$ is a solution corresponding to $|y_1,y_2\dots y_l\rangle$. Then we perform the measurement step, and the probability of obtaining some $s_{(y_1,y_2\dots y_l)}$ is $(\frac{1}{2^{n-1}})^l$. Likewise, there are r quantum states of $|y_1,y_2\dots y_l\rangle|s_{(y_1,y_2\dots y_l)}\rangle$ that contain the correct period $r$. The probability of getting the correct period is $\frac{r}{2^{(n-1)l}}$, which is the same as when measuring first. Parallel multiple Simon algorithms satisfy the principle of deferred measurement.

\section{Reanalysis of Grover-meets-Simon Algorithm\label{sec5}}

In this section, we reanalyze Grover-meets-Simon algorithm. According to section 4, after quantum Gaussian elimination algorithm, the probability of $U_{QGE}(|A_i\otimes0^{num}\rangle)$ does not change. That is $U_{QGE}(\sum_i\alpha_i(|A_i\rangle|0^{num}\rangle))=\sum_i\alpha_i(|A_i\rangle|s_i\rangle)$. $|A_i\rangle|s_i\rangle$ represents the solution $s_i$ corresponding to the matrix $|A_i\rangle$. Only considering the parallel Simon's algorithm, we can defer its measurement until the end. However, in the Grover-meets-Simon algorithm, it is equivalent to moving the measurement of all parallel Simon algorithms in the iterative process to the end of the entire Grover-meets-Simon algorithm. At this time, the feasibility of the Grover-meets-Simon algorithm needs to be reanalyzed.
% The initial sate that we need consider is formula (\ref{equ2}). And suppose that we can get the solution corresponding to the row simplest form matrix after quantum Gaussian elimination algorithm . We get that 

%  \begin{align*}
%  U_{QGE}|\varphi\rangle=&\frac{1}{\sqrt{2^m}}(\frac{1}{2^n})^l\sum_{i\in F_2^m,i\neq k}\sum_{x_1,x_2\dots x_l\in X_1}\sum_{y_1,y_2\dots y_l\perp s}(-1)^{x_1\cdot y_1}\dots(-1)^{x_l\cdot y_l}\\
% &\cdot|y_1,y_2\dots y_n, s\rangle|f(i,x_1)\rangle\dots|f(i,x_l)\rangle|i\rangle\\
% +&\frac{1}{\sqrt{2^m}}(\frac{1}{2^{n-1}})^l\sum_{x_1,x_2\dots x_l\in X_1}\sum_{y_1,y_2\dots y_l\perp s,s\neq 0}(-1)^{x_1\cdot y_1}\dots(-1)^{x_l\cdot y_l}\\
% &\cdot|y_1,y_2\dots y_n,s\rangle|f(k,x_1)\rangle\dots|f(k,x_l)\rangle|k\rangle
%  \end{align*}
% where$|y_1,y_2\dots y_n, s\rangle$ represents the solution $s$ corresponding to the matrix $|y_1,y_2\dots y_l\rangle$ 

Hence, we define classifier $U_g$ like \cite{leander2017grover} as follow: 

\begin{align*}
U_g:&F_2^m\times (F_2^{n})^l\rightarrow \{0,1\} \\
&(k',y_1,y_2\dots y_l)\rightarrow \{0,1\}
\end{align*}

(1) If $dim(y_1,y_2\dots y_l) \ne n-1$, output will be $0$. Otherwise, by quantum Gaussian elimination algorithm, we can get the solution $s$ corresponding to matrix $|y_1,y_2\dots y_l\rangle$. 

(2) Check $f(k',m)=f(k',m\oplus s)$ for an arbitrary provided plaintext $m$. If the identity holds, the output will be $1$, otherwise it will be $0$.

% Since registers of $|k'\rangle|y_1,y_2\dots y_l\rangle$ and registers of $|f(k',x_1)\rangle\dots|f(k',x_l)\rangle$ are entangled, the phase flip will affect the entire initial state after $U_g$.

Therefore, the operator $U_g$ can be define like:
\[
|\varphi'\rangle\rightarrow
\begin{cases}
-|\varphi'\rangle,\ \ if\ \ U_g|\varphi'\rangle=1\\
|\varphi'\rangle,\  \  if\ \ U_g|\varphi'\rangle=0
\end{cases}
\]
where $|\varphi'\rangle$ represents component of superposition state $|\varphi\rangle$.

% Because $|f(k',x_1),\dots ,f(k',x_l)\rangle$ is entangled with $|k',y_1,y_2,\dots,y_l$, the phase flip will affect the entire initial state components $|k'\rangle|y_1,y_2,\dots,y_l\rangle|f(k',k),\dots ,f(k,x_l)\rangle$ after $U_g$.
From formula (\ref{equ1}), we know that the the initial state of Grover-meets-Simon algorithm is is not a uniform superposition. At the same time there is an amplitude of oscillation. In \cite{biron1999generalized}, Biron et.al discussed the probability of success and query times of Grover's algorithm in arbitrary initial state. They depends only on the standard deviation of the initial amplitude distributions of the marked or unmarked states. And Hence, we need reanalyze the Grover-meets-Simon algorithm based on the initial amplitude. 

Next, we analyze the Grover-meets-Simon algorithm. The paper \cite{biron1999generalized} gives the maximum probability of success $P_{max}$ and the query times when the initial state is not a uniform superposition state.Before we discuss the maximum probability of success and query times, we first give some theorems.

\textbf{Theorem 3\cite{biron1999generalized}: } Suppose that the number of correct solutions is $r$ and the total number of databases searched is $N$. When $r/N\ll 1$ , the number of iterations $T$ of the search satisfies
\[
T=-\frac{1}{2}\frac{\overline{k(0)}}{\overline{l(0)}}+\frac{\pi}{4}\sqrt{N/r}-\frac{\pi}{24}\sqrt{r/N}+O(r/N)
\]
where $\overline{k(0)}$ represents the average of the initial probability amplitude of the marked states (correct solutions), and $\overline{l(0)}$ represents the average of the initial probability amplitude of the unmarked states. 

\textbf{Theorem 4:}Let $x_i\in F_2^n,i\in\{1,2\dots l\}$, when $y_i(i\in\{1,2\dots l\})$ are not all 0,
$$\sum_{x_1,x_2\dots x_l\in F_2^n}(-1)^{x_1\cdot y_1}\dots(-1)^{x_l\cdot y_l}=0$$

\textbf{Proof:} According to theorem in\cite{nielsen2002quantum}, when $x\in F_2^n$, we can get
\begin{equation}\label{eqeth4}
\begin{aligned}[b]
\sum_{x\in F_2^n}(-1)^{x\cdot y}=
\begin{cases}
0,\ \ y\neq 0\\
2^n,\ \ y=0
\end{cases}.
\end{aligned}
\end{equation}
Obviously,
\[
\sum_{x_1,x_2\dots x_l\in F_2^n}(-1)^{x_1\cdot y_1}\dots(-1)^{x_l\cdot y_l}=(\sum_{x_1\in F_2^n}(-1)^{x_1\cdot y_1})\cdot(\sum_{x_2\in F_2^n}(-1)^{x_2\cdot y_2})\cdot\dots\cdot(\sum_{x_l\in F_2^n}(-1)^{x_l\cdot y_l})
\]
Only when $y_i$ are all 0, the above formula is equal to $2^{nl}$. Theorem 4 is hold. \hfill$\qedsymbol$

Then, we discuss the maximum probability of success based on the initial
average amplitudes. 
According to the calculation method of the maximum probability of success in \cite{biron1999generalized}, we can get that $P_{max}=1-(N-r)\sigma_l^2$, where $\sigma_l^2=\frac{1}{N-r}\sum_{i=1}^{N-r}|l_i(0)-\overline{l(0)}|^2=\frac{1}{N-r}\sum_i|l_i(0)|^2-(\frac{1}{N-r}|\sum_il_i(0)|)^2$. According to formula (\ref{equ1}), we can know that 
\begin{equation}\label{eqel0}
\begin{aligned}[b]
l_i(0)^2=
\begin{cases}
\frac{1}{2^m}(\frac{1}{2^{2nl}}),\ \ k'\neq k\\
\frac{1}{2^m}(\frac{1}{2^{2(n-1)l}}),\ \ k=k
\end{cases}.
\nonumber
\end{aligned}
\end{equation}
Note that $k'=k$ but $dim(span(y_1,y_2\dots y_l))<n-1$
are unmarked states. In unmarked states, when $k'\neq k$, the number is $(2^m-1)2^{2nl}$ and when $k'=k$ ,the number is $2^{2(n-1)l}-r$. Hence, we can get
$$\sum_i|l_i(0)|^2=\frac{(2^m-1)2^{2nl}}{2^m2^{2nl}}+\frac{2^{2(n-1)l}-r}{2^m2^{2(n-1)l)}}\approx1-O(\frac{1}{2^m})$$
In addition, we show $\sum_il_i(0)$ as 
$$\sum_il_i(0)=\sum_{k'=k,dim(span(y_1\dots y_l))<n-1}l_i(0)+\sum_{k'\neq k}l_i(0)$$. 
Among them,
\begin{equation}
 \begin{split}
\sum_{k'=k,dim(span(y_1\dots y_l))<n-1}l_i(0)=\frac{1}{\sqrt{2^m}}(\frac{1}{\sqrt{2^{n-1}}})^l(\frac{1}{\sqrt{2^{n-1}}})^l\sum_{x_1,x_2\dots x_l\in X_1}\sum_{dim(span(y_1,y_2\dots y_l))<n-1}(-1)^{x_1\cdot y_1}\dots(-1)^{x_l\cdot y_l}\\
\nonumber
\end{split}
\end{equation} 
and
\begin{equation}
 \begin{split}
\sum_{k'\neq k}l_i(0)=\frac{1}{\sqrt{2^m}}(\frac{1}{\sqrt{2^n}})^l(\frac{1}{\sqrt{2^n}})^l\sum_{x_1,x_2\dots x_l\in F_2^n}\sum_{y_1,y_2\dots y_l\in F_2^n}(-1)^{x_1\cdot y_1}\dots(-1)^{x_l\cdot y_l}\\
\nonumber
\end{split}
\end{equation} 
When computing $\sum_{k'=k,dim(span(y_1\dots y_l))<n-1}l_i(0)$, we need to get the value of 
$$\sum_{x_1,x_2\dots x_l\in F_2^n}\sum_{y_1,y_2\dots y_l\in F_2^n}(-1)^{x_1\cdot y_1}\dots(-1)^{x_l\cdot y_l}$$.

According to equation (\ref{eqeth4}), we know that
\[
\sum_{x\in F_2^n}(-1)^{x\cdot y}=\sum_{x\in X_1}(-1)^{x\cdot y}+\sum_{x\oplus s\in F_2^n/ X_1}(-1)^{(x\oplus s)\cdot y}=
\begin{cases}
0,\ \ y\neq 0\\
2^n,\ \ y=0
\end{cases}
\]
Because of $y\perp s$, we can get$ \sum_{x\in F_2^n}(-1)^{x\cdot y}=2\sum_{x\in X_1}(-1)^{x\cdot y}$. Thus, we can get 
\begin{equation}\label{eqeK0}
\begin{aligned}[b]
\sum_{x\in X_1}(-1)^{x\cdot y}=
\begin{cases}
0,\ \ y\neq 0\\
2^{n-1},\ \ y=0
\end{cases}.
\end{aligned}
\end{equation}

According to the implementation of the $U_g$ operator, we know that $x\in X_1$, when $y_1,y_2\dots y_l$ are all 0, the value of $\sum_{k'=k,dim(span(y_1\dots y_l))<n-1}l_i(0)\neq 0$, $\sum_{k'\neq k}l_i(0)$ is the same. According to theorem 5 and formula (\ref{eqeK0}),
$$\sum_{k'=k,dim(span(y_1\dots y_l))<n-1}l_i(0)=\frac{2^{(n-1)l}}{\sqrt{2^m}2^{(n-1)l}}$$ and $$\sum_{k'\neq k}l_i(0)=\frac{(2^m-1)2^{nl}}{\sqrt{2^m}2^{nl}}$$

Hence,

$$(\frac{1}{N-r}|\sum_i l_i(0)|)^2=(\frac{1}{N-r}(\frac{(2^m-1)2^{nl}}{\sqrt{2^m}2^{nl}}+\frac{2^{(n-1)l}}{\sqrt{2^m}2^{(n-1)l}}))^2=(\frac{\sqrt{2^m}}{N-r})^2=\frac{2^m}{(N-r)^2}$$
By calculation, we can get the maximum probability of success
$$P_{max}=1-(N-r)\sigma_l^2\approx\frac{1}{2^m}-O(\frac{1}{2^m})+\frac{2^m}{N-r}\approx O(\frac{1}{2^m})+O(\frac{1}{2^{2nl}})$$

Through the maximum probability of success of the search of the whole space, we can see that the quantum state where the key $|k\rangle$ and $|y_1,y_2\dots y_l\rangle$ we want can not be found with a high probability in the whole space. It is also impossible to find the correct key $k$ with a high probability.
In addition, for comprehensive consideration, we analyze the number of iterations.

According to Theorem 3 and \cite{biham2002grover}, the optimal number of iterations has nothing to do with the initial state. To consider the minimum number of iterations, we consider the case of $n$ Simon's algorithm. The optimal qurey times are $T\~O(\sqrt{\frac{N}{r}})$.

Since the registers of multiple Simon algorithms are direct products, the matrices generated by them are random. That is, when $i$ is the correct key $k$, the matrices generated by them contain cases with rank less than or equal to $n-1$. By lemma 2, we can get the number of correct solutions.

\textbf{Lemma 3\label{lem3}:} Suppose that there is a vector space $V$, $y\in V$ and $y$ satisfies $y\in\{y|y\perp s=0,y,s\in F_2^n,s\neq 0\}$. Let $A$ be a $n\times n$ matrix and its row vectors belong to $V$, and $rank(A)=n-1$, then the number of the matrix $A$ is
\[
|\{A\}|=2^{\frac{(n-2)(n-1)}{2}}(2^n-1)\dots(2^{n-1}-1)\cdot(2^{n-2}-1)\dots(2-1)
\]

\textbf{Proof:} According to the relationship between the solution vector and the basic solution system, we can obtain the number of vectors in the vector space $V$ as $|V|=2^{n-1}$. To ensure $rank(A)=n-1$, we consider the number of possible values of each row. 
For the row where the first base is, $|\{y_1\}|=2^{n-1}-1(y_1\neq\vec{0})$. When considering the row where the second base is $y_2$, we need to ensure that it is linearly independent of $y_1$, so $|\{y_2\}|=2^{n-1}-2^1$. Similarly, for the row where the $i$-th row is, $|\{y_i\}|=2^{n-1}-2^{i-1}$. Thus,
\begin{equation}
 \begin{split}
|\{y_1\}|\cdot|\{y_2\}|\dots|\{y_{n-1}\}|&=(2^{n-1}-1)\cdot(2^{n-1}-2)\dots(2^{n-1}-2^{n-2})\\
&=2^{\frac{(n-2)(n-1)}{2}}(2^{n-1}-1)\cdot(2^{n-2}-1)\dots(2-1)
\nonumber
\end{split}
\end{equation}

For the $n\times n$ matrix $A$, one row is not a basis vector. If the row is the first row, it will be $\vec{0}$. If the row is the second row, it will be a vector with $y_1$ as the basis vector. Therefore, if the row is the $i-th$ row, it will be a vector with $y_1,y_2\dots y_{i-1}$ as a set of basis vector. Hence,
\begin{equation}
 \begin{split}
|\{A\}|&=|\{y_1\}|\cdot|\{y_2\}|\dots|\{y_{n-1}\}|\cdot(1+2^1+\dots+2^{n-1})\\
&=2^{\frac{(n-2)(n-1)}{2}}(2^n-1)\dots(2^{n-1}-1)\cdot(2^{n-2}-1)\dots(2-1)
\nonumber
\end{split}
\end{equation}

% The second case: There is a all-zero vector in the first $n-1$ rows. We randomly select a row from first $n-1$ rows for all-zero vector. The analysis for other lines is the same as the first case. Hence,
% \begin{equation}
%  \begin{split}
% |\{A_2\}|&=|\{y_1\}|\cdot|\{y_2\}|\dots|\{y_{n-1}\}|\cdot(n-1)\cdot (2^{n-1})^{l-n}\\
% &=(2^{n-1}-1)\cdot(2^{n-1}-2)\dots(2^{n-1}-2^{n-2})\cdot(n-1)\cdot (2^{n-1})^{l-n}\\
% &=2^{\frac{(n-1)(n-2)}{2}}(2^{n-1}-1)\cdot(2^{n-2}-1)\dots(2-1)\cdot(n-1)\cdot (2^{n-1})^{l-n}
% \nonumber
% \end{split}
% \end{equation}

% The number of matrix $A$ is
% $$|\{A\}|=|\{A_1\}|+|\{A_2\}|=2^{\frac{(n-1)(n-2)}{2}}(2^{n-1}-1)\cdot(2^{n-2}-1)\dots(2-1)\cdot(2^{n-1}+n-1)\cdot (2^{n-1})^{l-n}$$
% \rightline{$\qedsymbol$}
For convenience of calculation, we enlarge the value of $|\{A\}|$.

\begin{align*}
|\{A\}|=&2^{\frac{(n-1)(n-2)}{2}}(2^n-1)(2^{n-1}-1)\cdot(2^{n-2}-1)\dots(2-1)\\
&<2^{\frac{(n-1)(n-2)}{2}}2^n\cdot2^{n-1}\dots 2^2=2^{n(n-1)}
\end{align*}

According to lemma 2, We can compute the query times $T$:
\begin{equation}
 \begin{split}
\frac{N}{r}&=\frac{(2^m-1)2^{2n^2}+2^{2(n-1)n}}{(2^{n-1})^n|\{A\}|}\\
&>\frac{(2^m-1+2^{-2n})2^{2n^2}}{2^{2n(n-1)}}\\
&>2^{m+2n}-2^{2n}
\nonumber
\end{split}
\end{equation}
Hence, the query times $T>O(\sqrt{(2^{m}-1)2^{2n}})$.
The query times of this algorithm are much higher than the query times of searching two keys directly using the quantum exhaustive search, which is $T\sim O(\sqrt{2^{m+n}})$. 

% Therefore, after analyzing solution of quantum system of linear equations, we re-analyzed the query times of Grover-meets-Simon algorithm. The entanglement makes that Grover-meets-Simon algorithm is not an effective attack algorithm. Although this algorithm can be implemented in a quantum environment, its efficiency still needs to be improved.

When considering the amplitude of the initial state, the initial amplitude is not a uniform superposition state, and there is an oscillation in the amplitude. After analyzing feasibility of Grover-meets-Simon algorithm, we can know that Grover-meets-Simon algorithm cannot obtain the correct key through the delay measurement when moving the measurement of all parallel Simon algorithms in the iterative process to the end of the entire algorithm.

\section{Conclusions}

For the discussion of many quantum algorithms, it is necessary to consider the actual operation in the quantum environment. For the combination of different quantum algorithms, especially the combination with Grover's algorithm, the measurement problem needs to be discussed emphatically. In this paper, we reanalyze the existing Grover-meets-Simon algorithm from the perspective of the principle of deferred measurement.

In the original algorithm, using the general deferred measurement principle of quantum computation, the measurement is placed at the end of the algorithm, so we need to consider the intermediate operation of the entire algorithm. The output of multiple Simon registers in the Grover-meets-Simon algorithm is the superposition state of the matrix when not measuring. Therefore, we need discuss the situation of the solution of quantum linear systems of equations. First, we give the definition of the quantum linear systems of equations and discuss the problems encountered when using the Gaussian elimination algorithm to solve the quantum linear systems of equations. According to the analysis, quantum Gaussian elimination algorithm can realize that $U_{QGE}(\sum_i\alpha_i(|A_i\rangle|0^{num}\rangle))=\sum_i\alpha_i(|A_i\rangle|s_i\rangle)$. And based on quantum Gaussian elimination algorithm, only considering the parallel Simon's algorithm, we can defer its measurement until the end. However,Grover-meets-Simon algorithm put every measurement in Simon's algorithm involves multiple iterations of the Simon's algorithm, directly putting all Simon's algorithm measurements at the end will have an impact. We re-analyzed the maximum probability of success and the query times of Grover-meets-Simon algorithm. We can not find correct key $k$ with high probability. In this case, Grover-meets-Simon algorithm is not an effective attack algorithm. Therefore, it is not feasible to directly put the measurement process of the Simon algorithm in the iterative process at the end of Grover-meets-Simon algorithm.

In the following work, on the one hand, inspired by this paper, we may try to discuss the feasibility of other cryptanalysis algorithms for practical quantum algorithm construction. On the other hand, the discussion of quantum linear systems of equations is based on the quantum Gaussian elimination algorithm. For the solution of quantum linear systems of equations, we can start with the fundamental idea of classical problems and consider their general feasibility.

\section*{Appendix A}
\subsection*{A.1}
In addition to the quantum Gaussian elimination algorithm, we consider the quantum implementation of another classical algorithm for solving linear systems of equations (quantum Gauss Jordan elimination algorithm).

\begin{algorithm}[H]
		\caption{\textbf{Quantum Gauss Jordan elimination algorithm}}
		\label{Alg:QGJEA}
		\begin{algorithmic}[1]
			\Require 
		    $n(n+1)$ qubits to store the augmented matrix value of $[A|b]$:$|a_{1,1},a_{1,2}\dots,a_{1,n},a_{2,1},\dots a_{2,n},\dots a_{n,n},b_1,b_2\dots b_n\rangle$ $O(n^2)$ auxiliary qubits to ensure that the algorithm
			\Ensure the value of a vector $\vec{x}$ such at $A\vec{x}=\vec{b}$
			\State \textbf{for} $j\leftarrow 1$ $\textbf{to}$ $n-1$ 
			\State \ \ \ \ $\textbf{for} \ \ \ i\leftarrow j+1 $     $\textbf{to}$ $n$
			\State \ \ \ \ \ \ \ \   $\textbf{if}$  $a_{j,j}=0$   $\textbf{do}$  $a_{j,:}=a_{j,:}\bigoplus a_{i,:},b_j=b_j\oplus b_i$
			\State \ \ \ \ $\textbf{for} \ \ \ k\leftarrow 1 $     $\textbf{to}$ $n$
			\State \ \ \ \ \ \ \ \   $\textbf{if}$  $(k\neq j)\bigwedge(a_{k,j}=1)$   $\textbf{do}$  $a_{k,:}=a_{j,:}\bigoplus a_{k,:},b_k=b_k\oplus b_j$
			\State \textbf{for}\ \ $r\leftarrow 1$ \textbf{to} $n-1$
			\State \ \ \ \ \textbf{if} $a_{r,n}=1$ \textbf{do} $a_{r,n}=a_{n,n}\oplus a_{r,n}$ , $b_r=b_r\oplus b_n$
			\State \textbf{return} $x=b$
		\end{algorithmic}
\end{algorithm}

Quantum Gauss Jordan elimination algorithm is similar to quantum Gaussian elimination algorithm, they both use the same solution theorem of equations. But quantum Gauss Jordan elimination algorithm omits the process of back substitution by change augmented matrix into row simplest form matrix.The calculation process is similar to the analysis of quantum Gaussian elimination algorithm.

\noindent\textbf{Quantum circuit of quantum Gauss Jordan elimination algorithm:}

Now, we construct corresponding quantum circuits for quantum Gauss Jordan elimination algorithm. In addition, we estimate the quantum resources required for the quantum circuit. 

According to algorithm \ref{Alg:QGJEA}, the quantum Gauss Jordan elimination algorithm does not include the back substitution process.

\begin{figure}[h!]
\centerline{  \includegraphics[width=8cm]{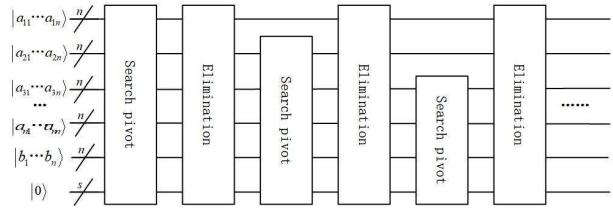}}
  \caption{Quantum Gauss Jordan elimination algorithm}
  \label{QGJE}
\end{figure}

Compared with quantum Gaussian elimination algorithm, the searching pivot process of quantum Gauss Jordan elimination algorithm remains unchanged, but all elements in each column except the pivot are eliminated to 0 during elimination, like figure \ref{QGJE}. When we eliminate $i$-th column, let function $f_3$ be:
\[
f_3(a_{i,j},a_{k,j})=(a_{i,j}, a_{i,j}\oplus a_{k,j}) \ \ \forall j\ge i\ \ and\ \ k\neq i.
\]

Because the searching pivot process of quantum Gauss Jordan elimination algorithm is same. To search the pivot of $i $ th column, we need $n-i$ CNOT gates, $(n-i)(n-i+2)$ Toffoli gates and $n-i$ auxiliary qubits. For elimination process, we need $2(n-1)$ CNOT gates, $(n-1)(n-i+1)$ Toffoli gates and $n-1$ auxiliary qubits. Hence, Quantum Gauss Jordan elimination algorithm needs $5n^3+\frac{11}{2}n^2-\frac{21}{2}n$ CNOT gates. And T-depth is $\frac{7n(n-1)(5n+8)}{6}$, auxiliary qubits are $\frac{3n(n-1)}{2}$.

\begin{figure}[h!]
 \centerline{  \includegraphics[width=10cm]{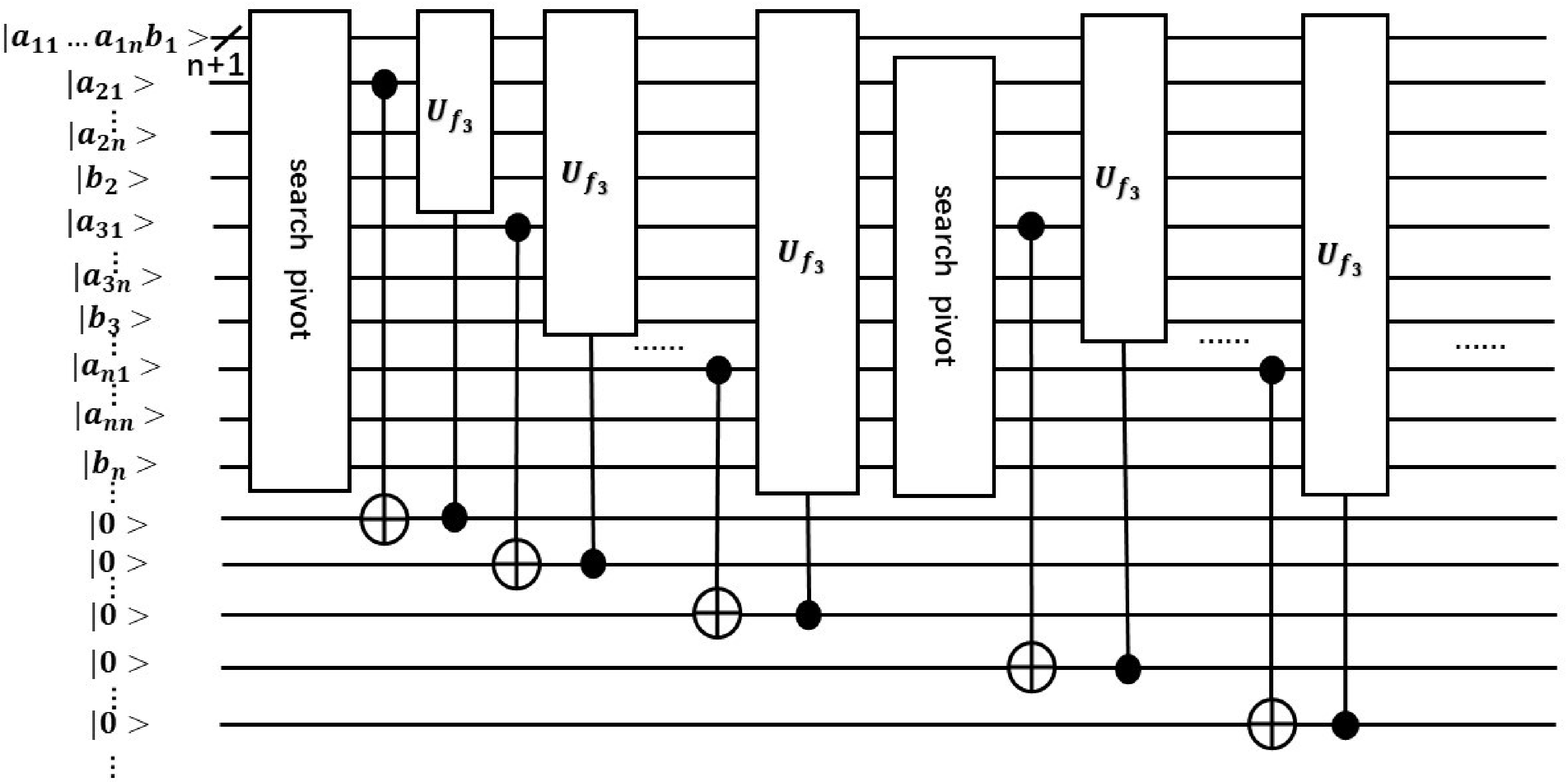}}
  \caption{Quantum Gauss Jordan elimination algorithm}
  \label{QGJE E}
\end{figure}

\subsection*{A.2}
Here, we discuss based on the quantum Gauss Jordan elimination algorithm. Algorithm \ref{Alg:IQGJE1} give an example, which the input matrix is a $n\times n$ matrix. See Appendix A.3 for complete algorithm of $m\times n$ matrix.

\begin{algorithm}[H]
		\caption{\textbf{row simplest form matrix computation}}
		\label{Alg:IQGJE1}
		\begin{algorithmic}[1]
			\Require 
		    $n\times n$ bit to store the matrix value of $A$ 
			\Ensure Row simplest form matrix $A'$ corresponding to matrix $A$
			\State \textbf{for} $i\leftarrow 1$ $\textbf{to}$ $n-1$
			\State \ \ \ \  $\textbf{for} \ \ \ j\leftarrow i+1 $     $\textbf{to}$ $n$
			\State \ \ \ \ \ \ \ \   $\textbf{if}$  $a_{i,i}=0$   $\textbf{do}$  $a_{i,:}=a_{i,:}\bigoplus a_{j,:}$
			\State \ \ \ \  $\textbf{for} \ \ \ k\leftarrow 1 $     $\textbf{to}$ $n$
			\State \ \ \ \ \ \ \ \    $\textbf{if}$  $(k\neq i)\wedge(a_{k,i}=1)$   $\textbf{do}$ $a_{k,:}=a_{i,:}\bigoplus a_{k,:}$
	     	\State \ \ \ \ \ \ \ \  $\textbf{for} \ \ \ p\leftarrow i$     $\textbf{to}$ $n-1$
	     	\State \ \ \ \ \ \ \ \ \ \ \ \    $\textbf{if}$  $\forall l\in[i,p] \ \ a_{i,l}=0$   $\textbf{do}$
	     	\State \ \ \ \ \ \ \ \ \ \ \ \ \ \ \  $\textbf{for} \ \ \ q\leftarrow i+1 $     $\textbf{to}$ $n$
	       	\State \ \ \ \ \ \ \ \ \ \ \ \ \ \ \ \ \ \ \textbf{if} $a_{i,p+1}=0$ \textbf{do} $a_{i,:}=a_{i,:}\bigoplus a_{q,:}$
	       	\State \ \ \ \ \ \ \ \ \ \ \ \ \ \ \  $\textbf{for} \ \ \ r\leftarrow 1 $     $\textbf{to}$ $m$
	       	\State \ \ \ \ \ \ \ \ \ \ \ \ \ \ \ \ \ \ $\textbf{if}$  $(r\neq i)\wedge(a_{r,p+1}=1)$   $\textbf{do}$ $a_{r,:}=a_{q,:}\bigoplus a_{r,:}$
	       	\State $\textbf{for} \ \ k\leftarrow 1 $     $\textbf{to}$ $n-1$
			\State \ \ \ \ $\textbf{if}$  $a_{k,n}=1$   $\textbf{do}$ $a_{k,n}=a_{n,n}\bigoplus a_{k,n}$
	       	\State \textbf{return $A'$}
		\end{algorithmic}
\end{algorithm}

Due to the characteristics of quantum computer, we can not get the intermediate results in the operation process. Algorithm \ref{Alg:IQGJE1} uses the property of row simplest form matrix ( according to definition 2 ) to ensure the universality and continuity of quantum operation. The rank of the matrix is equal to the number of non-zero rows of its row simplest form matrix. Therefore, the number of non-zero rows of an row simplest form matrix is less than or equal to $n$. We only need to search the elements of the upper triangle to determine the position of the pivot. 
Considering another characteristic of row simplest form matrix,  if $a_{i,j}=1$, $\forall k>i, a_{k,j}=0$ and  $\forall l<j, a_{i,l}=0$, then $a_{i,j}$ is pivot. Using multi-controlled gates, we control the exchange and elimination of rows by this condition.

Quantum circuit of searching pivot process and elimination process in the improved quantum Gauss Jordan elimination algorithm is same as quantum Gaussian Jordan elimination algorithm. But since we are not sure whether a column belong to the maximal linearly independent system, we need to judge each element of a row.
Consider the operation of the $i$ th row, we find the pivot through multi-controlled gates. Only when $\forall k\in [i,j-1],\ \ a_{i,k}=0$ and $a_{i,j}=1$, $a_{i,j}$ is pivot. For the $i$ th row, we need $O(n)$ multi-controlled gates(we use an auxiliary qubit to store condition $\forall k\in [i,j-1],\ \ a_{i,k}=0$). Decomposing multi-controlled gates, $O(n^2)$ toffoli gates can be get. However, due to the addition of a condition to control the process of pivot search and elimination. The CNOT gates in the quantum Gauss Jordan elimination algorithm have become 2-fold CNOT gates, and the Toffoli gates have become 3-fold CNOT gates. Therefore, we need $O(n)$ CNOT gates and $O(n^3)$ Toffoli gates. Then input a random matrix in improved quantum Gauss Jordan elimination algorithm, we need $O(n^2)$ CNOT gates and $O(n^4)$ Toffoli gates.

\begin{figure}[h!]
\centerline{\includegraphics[width=10cm]{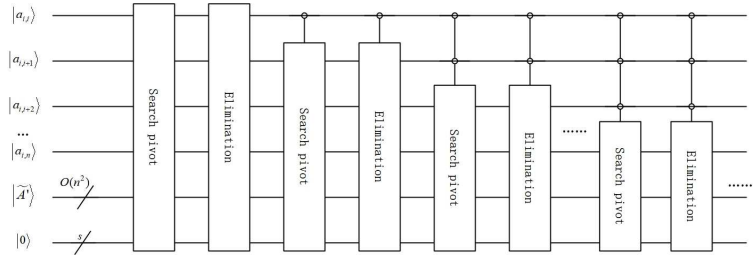}}  
  \caption{Improved quantum Gauss Jordan elimination algorithm (the i th row pivot searching) ($|\widetilde{A’}\rangle$ represents that matrix $|A\rangle$ removes the qubits of ($|a_{i,i}\rangle,|a_{i,i+1}\rangle\dots|a_{i,n}\rangle$), but not all qubits in $|\widetilde{A’}\rangle$ participate in the operation)}
\end{figure}

\subsection*{A.3}

The algorithm \ref{Alg:IQGJE} is the complete situation of algorithm \ref{Alg:IQGJE1} . In this algorithm, the input matrix is random matrix.

\begin{algorithm}[H]
		\caption{\textbf{row simplest form matrix computation(Full version)}}
		\label{Alg:IQGJE}
		\begin{algorithmic}[1]
			\Require 
		    $mn$ bit to store the matrix value of $A$ 
			\Ensure Row simplest form matrix $A'$ corresponding to matrix $A$
			\State \textbf{if} $n<m$ \textbf{do}
			\State \ \ \ \ \textbf{for} $i\leftarrow 1$ $\textbf{to}$ $n-1$
			\State \ \ \ \ \ \ \ \ $\textbf{for} \ \ \ j\leftarrow i+1 $     $\textbf{to}$ $m$
			\State \ \ \ \ \ \ \ \ \ \ \ \   $\textbf{if}$  $a_{i,i}=0$   $\textbf{do}$  $a_{i,:}=a_{i,:}\bigoplus a_{j,:}$
			\State \ \ \ \ \ \ \ \  $\textbf{for} \ \ \ k\leftarrow 1 $     $\textbf{to}$ $m$
			\State \ \ \ \ \ \ \ \ \ \ \ \   $\textbf{if}$  $(k\neq i)\wedge(a_{k,i}=1)$   $\textbf{do}$ $a_{k,:}=a_{i,:}\bigoplus a_{k,:}$
	     	\State \ \ \ \ \ \ \ \  $\textbf{for} \ \ \ p\leftarrow i$     $\textbf{to}$ $n-1$
	     	\State \ \ \ \ \ \ \ \ \ \ \ \    $\textbf{if}$  $\forall l\in[i,p] \ \ a_{i,l}=0$   $\textbf{do}$
	     	\State \ \ \ \ \ \ \ \ \ \ \ \ \ \ \  $\textbf{for} \ \ \ q\leftarrow i+1 $     $\textbf{to}$ $m$
	       	\State \ \ \ \ \ \ \ \ \ \ \ \ \ \ \ \ \ \ \textbf{if} $a_{i,p+1}=0$ \textbf{do} $a_{i,:}=a_{i,:}\bigoplus a_{q,:}$
	       	\State \ \ \ \ \ \ \ \ \ \ \ \ \ \ \  $\textbf{for} \ \ \ r\leftarrow 1 $     $\textbf{to}$ $m$
	       	\State \ \ \ \ \ \ \ \ \ \ \ \ \ \ \ \ \ \ $\textbf{if}$  $(r\neq i)\wedge(a_{r,p+1}=1)$   $\textbf{do}$ $a_{r,:}=a_{q,:}\bigoplus a_{r,:}$
	        \State \ \ \ \  $\textbf{for} \ \ j\leftarrow n+1 $     $\textbf{to}$ $m$
			\State \ \ \ \ \ \ \ \ \ \ \ \   $\textbf{if}$  $a_{n,n}=0$   $\textbf{do}$  $a_{n,n}=a_{n,n}\bigoplus a_{j,n}$
			\State \ \ \ \ \ \ \ \  $\textbf{for} \ \ \ k\leftarrow 1 $     $\textbf{to}$ $m$
			\State \ \ \ \ \ \ \ \ \ \ \ \   $\textbf{if}$  $(k\neq n)\wedge(a_{k,n}=1)$   $\textbf{do}$ $a_{k,n}=a_{n,n}\bigoplus a_{k,n}$
	       	\State \textbf{else} $n\ge m$ \textbf{do}
	       	\State \ \ \ \ \textbf{for} $i\leftarrow 1$ $\textbf{to}$ $m-1$
			\State \ \ \ \ \ \ \ \ repeat step 3-12	    
			\State \ \ \ \ \textbf{for} $p\leftarrow m-1$ $\textbf{to}$ $n$
	     	\State \ \ \ \ \ \ \ \  $\textbf{if}$  $(\forall l\in[m-1,p] \ , \ a_{m,l}=0)\wedge (a_{m,p+1}=1)$   $\textbf{do}$
	       	\State \ \ \ \ \ \ \ \  \ \ \ \  $\textbf{for} \ \ \ r\leftarrow 1 $     $\textbf{to}$ $m-1$
	       	\State \ \ \ \ \ \ \ \ \ \ \ \ \ \ \ \ \ $\textbf{if}$  $a_{r,p+1}=1$   $\textbf{do}$ $a_{r,:}=a_{m,:}\bigoplus a_{r,:}$			
	       	\State \textbf{return $A'$}
		\end{algorithmic}
\end{algorithm}

\section*{Data Availability}

The data used are available within the article.

\section*{Conflicts of Interest}

The author(s) declare(s) that there is no conflict of interest regarding the publication of this paper.

\section*{Funding Statement}

This work was supported by National Natural Science Foundation of China (Grant No. 61672517), National Natural Science Foundation of China (Key Program, Grant No. 61732021).

\bibliographystyle{unsrt}
\bibliography{ref} % see references.bib for bibliography management

\end{document}